\newif\ifOneCol

\ifOneCol
	\documentclass[draftcls,onecolumn,11pt]{IEEEtran}
\else
	\documentclass[twocolumn,10pt]{IEEEtran}
\fi

\usepackage[cmex10]{amsmath}
\usepackage{amsfonts}
\usepackage{verbatim}
\usepackage{multirow}

\usepackage{cite}

\ifCLASSINFOpdf
  \usepackage[pdftex]{graphicx}
  \graphicspath{{graphics/}}
  \DeclareGraphicsExtensions{.pdf,.jpeg,.png}
\else
  \usepackage[dvips]{graphicx}
  \graphicspath{{graphics/}}
  \DeclareGraphicsExtensions{.eps}
\fi

\usepackage{array}

\usepackage{mdwmath}
\usepackage{mdwtab}
\DeclareMathOperator{\erf}{erf}

\DeclareMathOperator{\erfc}{erfc}

 \newcommand{\sgn}{\operatorname{sgn}}

\begin{document}
\bibliographystyle{IEEEtran}
%
\title{A Unifying Model for External Noise Sources and \ISI\,
in Diffusive Molecular Communication}

\author{Adam Noel, \IEEEmembership{Student Member, IEEE}, Karen C. Cheung, and
	Robert Schober,
	\IEEEmembership{Fellow, IEEE}
\thanks{Manuscript received October 21, 2013; revised April 21, 2014; accepted
		June 3, 2014. This work was supported by the Natural
		Sciences and Engineering Research Council of Canada, and a Walter C. Sumner
		Memorial Fellowship. Computing resources were provided by WestGrid and
		Compute/Calcul Canada.}
\thanks{The authors are with the Department of Electrical and Computer Engineering, University of British Columbia, Vancouver, BC, Canada,	V6T 1Z4 (email: \{adamn, kcheung, rschober\}@ece.ubc.ca). R. Schober is also with the Institute for Digital Communication, Friedrich-Alexander-Universit\"{a}t Erlangen-N\"{u}rnberg (FAU), Erlangen, Germany (email: schober@lnt.de).}}


\newcommand{\dbydt}[1]{\frac{d#1}{dt}}
\newcommand{\pbypx}[2]{\frac{\partial #1}{\partial #2}}
\newcommand{\psbypxs}[2]{\frac{\partial^2 #1}{\partial {#2}^2}}
\newcommand{\dbydtc}[1]{\dbydt{\conc{#1}}}
\newcommand{\thev}{\theta_v}
\newcommand{\thevi}[1]{\theta_{v#1}}
\newcommand{\theh}{\theta_h}
\newcommand{\thehi}[1]{\theta_{h#1}}
\newcommand{\x}{x}
\newcommand{\y}{y}
\newcommand{\z}{z}
\newcommand{\rad}[1]{\vec{r}_{#1}}
\newcommand{\radmag}[1]{|\rad{#1}|}

\newcommand{\kth}[1]{k_{#1}}
\newcommand{\km}{K_M}
\newcommand{\vm}{v_{max}}
\newcommand{\conc}[1]{[#1]}
\newcommand{\conco}[1]{[#1]_0}
\newcommand{\C}{C}
\newcommand{\Cx}[1]{C_{#1}}
\newcommand{\CxFun}[3]{C_{#1}(#2,#3)}
\newcommand{\Cobs}{C_{obs}}
\newcommand{\Nobs}{{\Nx{\A}}_{obs}}
\newcommand{\Nobst}[1]{\Nobs\!\left(#1\right)}
\newcommand{\Nobsn}[1]{\Nobs\left[#1\right]}
\newcommand{\Nobsavgt}{\overline{{\Nx{\A}}_{obs}}(t)}
\newcommand{\Nobsavg}[1]{\overline{{\Nx{\A}}_{obs}}\left(#1\right)}
\newcommand{\Nobsavgmax}{\overline{{\Nx{\A}}_{max}}}
\newcommand{\Nnoisetavg}[1]{\overline{{\Nx{\A}}_{n}}\left(#1\right)}
\newcommand{\DMLSNntavg}[1]{\overline{{\Nx{\DMLSA}}_{n}^\star}\left(#1\right)}
\newcommand{\DMLSNnavg}{\overline{{\Nx{\DMLSA}}_{n}^\star}}
\newcommand{\DMLSNxavg}[1]{\overline{{\Nx{\DMLSA}}_{#1}^\star}}
\newcommand{\DMLSNxtavg}[2]{\overline{{\Nx{\DMLSA}}_{#2}^\star}\left(#1\right)}
\newcommand{\DMLSNxt}[2]{{\Nx{\DMLSA}}_{#2}^\star\left(#1\right)}
\newcommand{\DMLSNx}[1]{{\Nx{\DMLSA}}_{#1}^\star}
\newcommand{\Nnoiset}[1]{{\Nx{\A}}_{n}\left(#1\right)}
\newcommand{\Ntxt}[1]{{\Nx{\A}}_{TX}\left(#1\right)}
\newcommand{\Ntxtavg}[1]{\overline{{\Nx{\A}}_{TX}}\left(#1\right)}
\newcommand{\DMLSNtxtavg}[1]{\overline{{\Nx{\DMLSA}}_{tx}^\star}\left(#1\right)}
\newcommand{\DMLSNtxt}[1]{{\Nx{\DMLSA}_{tx}^\star}\left(#1\right)}
\newcommand{\DMLSNtxtavgU}[2]{\overline{{\Nx{\DMLSA}}_{tx,#2}^\star}\left(#1\right)}
\newcommand{\DMLSNtxtU}[2]{{\Nx{\DMLSA}_{tx,#2}^\star}\left(#1\right)}
\newcommand{\Cgen}{C_A(r, t)}
\newcommand{\radbind}{r_B}

\newcommand{\M}{M}
\newcommand{\smM}{m}
\newcommand{\A}{A}
\newcommand{\X}{S}
\newcommand{\vx}[1]{v_{#1}}
\newcommand{\vxvec}[1]{\vec{v}_{#1}}
\newcommand{\Pec}[1]{v^\star_{#1}}
\newcommand{\Pecper}{\Pec{\perp}}
\newcommand{\Pecpara}{\Pec{\scriptscriptstyle\parallel}}
\newcommand{\metre}{\textnormal{m}}
\newcommand{\second}{\textnormal{s}}
\newcommand{\molecule}{\textnormal{molecule}}
\newcommand{\bound}{\textnormal{bound}}
\newcommand{\argmax}{\operatornamewithlimits{argmax}}
\newcommand{\Dx}[1]{D_{#1}}
\newcommand{\Nx}[1]{N_{#1}}
\newcommand{\NAx}[1]{N_{\A_{#1}}}
\newcommand{\NAxavgt}[2]{\overline{\NAx{#2}}\left(#1\right)}
\newcommand{\NAxavg}[1]{\overline{\NAx{#1}}}
\newcommand{\Nemit}{\Nx{{\A_{EM}}}}
\newcommand{\NemitU}[1]{\Nx{{\A_{EM,#1}}}}
\newcommand{\Ngen}{\Nx{{\A_{gen}}}}
\newcommand{\Ngent}[1]{\Nx{{\A_{gen}}}\left(#1\right)}
\newcommand{\Ngenavgt}[1]{\overline{\Nx{{\A_{gen}}}}\left(#1\right)}
\newcommand{\DMLSNgent}[1]{\Nx{{\DMLSA_{gen}}}^\star\left(#1\right)}
\newcommand{\DMLSNgenavgt}[1]{\overline{\Nx{{\DMLSA_{gen}}}^\star}\left(#1\right)}
\newcommand{\Da}{D_\A}
\newcommand{\En}{E}
\newcommand{\en}{e}
\newcommand{\Ne}{\Nx{\En}}
\newcommand{\De}{D_\En}
\newcommand{\EA}{EA}
\newcommand{\ea}{ea}
\newcommand{\Nint}{\Nx{\EA}}
\newcommand{\Di}{D_{\EA}}
\newcommand{\Etot}{\En_{Tot}}
\newcommand{\stepl}{r_{rms}}
\newcommand{\AP}{A_P}
\newcommand{\Ri}[1]{R_{#1}}
\newcommand{\ro}{r_0}
\newcommand{\rone}{r_1}
\newcommand{\visc}{\eta}
\newcommand{\bolt}{\kth{B}}
\newcommand{\temp}{T}
\newcommand{\T}{T_{int}}
\newcommand{\TU}[1]{T_{int,#1}}
\newcommand{\Vobs}{V_{obs}}
\newcommand{\robs}{r_{obs}}
\newcommand{\Ve}{V_{enz}}
\newcommand{\tint}{\delta t}
\newcommand{\tmax}{t_{max}}
\newcommand{\Cobsfrac}{\alpha}
\newcommand{\dist}{L}
\newcommand{\DMLSA}{a}
\newcommand{\DMLSt}[1]{t_{#1}^\star}
\newcommand{\DMLST}{\T^\star}
\newcommand{\DMLSTU}[1]{\TU{#1}^\star}
\newcommand{\DMLStau}{\tau}
\newcommand{\DMLSx}{x^\star}
\newcommand{\DMLSy}{y^\star}
\newcommand{\DMLSz}{z^\star}
\newcommand{\DMLSr}[1]{r_{#1}^\star}
\newcommand{\DMLSrad}[1]{\rad{#1}^\star}
\newcommand{\DMLSradmag}[1]{|\DMLSrad{#1}|}
\newcommand{\DMLSk}[1]{\kth{#1}^\star}
\newcommand{\DMLSC}[1]{\Cx{#1}^\star}
\newcommand{\DMLSCxFun}[3]{{\DMLSC{#1}}(#2,#3)}
\newcommand{\DMLSc}[1]{\gamma_{#1}}
\newcommand{\DMLSV}{\Vobs^\star}
\newcommand{\DMLSNemit}{{\Nx{{{\DMLSA}_{EM}}}^\star}}
\newcommand{\DMLSNemitU}[1]{{\Nx{{{\DMLSA}_{EM,#1}}}^\star}}
\newcommand{\DMLSNA}{\overline{{\Nx{\DMLSA}}_{obs}^\star}(t)}
\newcommand{\DMLSNAb}{\overline{{\Nx{\DMLSA}}_{obs}^\star}(\DMLSt{B})}
\newcommand{\DMLSNAmax}{{\overline{{\Nx{\DMLSA}}_{max}^\star}}}
\newcommand{\DMLStmax}[1]{{\DMLSt{#1}}_{,max}}
\newcommand{\DMLSdim}{\mathcal{D}}
\newcommand{\DMLSthreshInt}{\alpha^\star}
\newcommand{\DMLSv}[1]{v^\star_{#1}}
\newcommand{\DMLSvxvec}[1]{\vec{v}^\star_{#1}}

\newcommand{\data}[1]{W\left[#1\right]}
\newcommand{\dataU}[2]{W_{#2}\left[#1\right]}
\newcommand{\dataSeq}{\mathbf{W}}
\newcommand{\dataSeqU}[1]{\mathbf{W_{#1}}}
\newcommand{\dataSet}{\mathcal{W}}
\newcommand{\dataObs}[1]{\hat{W}\left[#1\right]}
\newcommand{\numX}[2]{n_{#1}\left(#2\right)}
\newcommand{\thresh}{\xi}
\newcommand{\DMLSthresh}{\xi^\star}
\newcommand{\poissBar}{\Big|_\textnormal{Poiss}}
\newcommand{\gaussBar}{\Big|_\textnormal{Gauss}}
\newcommand{\eqBar}[2]{\Big|_{#1 = #2}}
\newcommand{\Pobs}{P_{obs}}
\newcommand{\Pobsx}[1]{P_{obs}\left(#1\right)}
\newcommand{\Pone}{P_1}
\newcommand{\Pzero}{P_0}
\newcommand{\Pe}[1]{P_e\left[#1\right]}
\newcommand{\Peavg}[1]{\overline{P}_e\left[#1\right]}
\newcommand{\threshInterval}{\alpha}
\newcommand{\pstay}[1]{P_{stay}\left(#1\right)}
\newcommand{\pleave}[1]{P_{leave}\left(#1\right)}
\newcommand{\parrive}[1]{P_{arr}\left(#1\right)}

\newcommand{\VAmem}{F}
\newcommand{\VAstate}{f}
\newcommand{\VAdataObs}[2]{\hat{W}_{\VAstate_{#2}}\left[#1\right]}
\newcommand{\VAcurLL}[2]{\Phi_{\VAstate_{#2}}\left[#1\right]}
\newcommand{\VAcumLL}[2]{L_{\VAstate_{#2}}\left[#1\right]}

\newcommand{\weight}[1]{w_{#1}}

\newcommand{\fof}[1]{f\left(#1\right)}
\newcommand{\floor}[1]{\lfloor#1\rfloor}
\newcommand{\lam}[1]{W\left(#1\right)}
\newcommand{\EXP}[1]{\exp\left(#1\right)}
\newcommand{\ERF}[1]{\erf\left(#1\right)}
\newcommand{\ERFC}[1]{\erfc\left(#1\right)}
\newcommand{\SIN}[1]{\sin\left(#1\right)}
\newcommand{\SINH}[1]{\sinh\left(#1\right)}
\newcommand{\COS}[1]{\cos\left(#1\right)}
\newcommand{\COSH}[1]{\cosh\left(#1\right)}
\newcommand{\Ix}[2]{I_{#1}\!\left(#2\right)}
\newcommand{\Jx}[2]{J_{#1}\!\left(#2\right)}
\newcommand{\E}[1]{E\left[#1\right]}
\newcommand{\GamFcn}[1]{\Gamma\!\left(#1\right)}
\newcommand{\mean}[1]{\mu_{#1}}
\newcommand{\var}[1]{\sigma_{#1}^2}

\newcommand{\B}[1]{B_{#1}}
\newcommand{\w}{w}
\newcommand{\n}{n}
\newcommand{\gx}[1]{g\left(#1\right)}
\newcommand{\hx}[1]{h\left(#1\right)}
\newcommand{\tx}[1]{t\left(#1\right)}
\newcommand{\ux}[1]{u\left(#1\right)}
\newcommand{\deltObs}{t_{o}}
\newcommand{\sx}[1]{s_{#1}}
\newcommand{\DMLSs}[1]{s_{#1}^\star}
\newcommand{\U}{U}

\newcommand{\new}[1]{\textbf{#1}}
\newcommand{\ISI}{ISI}
\newcommand{\DDFSE}{DDFSE}
\newcommand{\PDF}{PDF}
\newcommand{\CDF}{CDF}
\newcommand{\AWGN}{AWGN}
\newcommand{\UCA}{UCA}

\newtheorem{theorem}{Theorem}
\newtheorem{remark}{Remark}


\newcommand{\tableOverview}[5]{
	\begin{table}[#1]
	\centering
	\caption{Description of the terms in (\ref{EQ13_05_28_obs}).}
	{\renewcommand{\arraystretch}{1.2}
		\begin{tabular}{|c|p{#3}|c|c|p{#5}|}
		\hline
		\parbox[t]{#2}{Component\\ of $\DMLSNxt{\DMLSt{}}{obs}$} &
		Source Type & Section &
		\parbox[t]{#4}{Detailed\\ Form} &
		\parbox[t]{#5}{Emitting\\ Continuously?} \\ \hline
		\multirow{2}{*}{$\DMLSNxt{\DMLSt{}}{u}$} &
		Random Noise & \ref{sec_noise} & (\ref{APR12_42_DMLS}) & Yes\\ \cline{2-5}
		 & Interfering Transmitter & \ref{sec_mui} &
		(\ref{EQ13_08_02}) & Approximation in (\ref{EQ13_08_05}) \\ \hline
		$\DMLSNxt{\DMLSt{}}{1}$ & Intended Transmitter & \ref{sec_isi} &
		(\ref{EQ13_08_08_DMLS}) & Approximation of old ISI in
		(\ref{EQ13_08_09})-(\ref{EQ13_05_18_isi}) \\ \hline
		\end{tabular}
	}
	\label{table_overview}
	\end{table}
}

\newcommand{\tableNoise}[1]{
	\begin{table}[#1]
	\centering
	\caption{Summary of the equations for the impact of an external noise
	source and the conditions under which they can be used. By $\DMLSx_n =$
	``Far'', we mean that the \UCA\, is applied.}
	{\renewcommand{\arraystretch}{1.2}
		\begin{tabular}{|c|c|c|c|c|c|c|}
		\hline
		Eq. & Closed Form? & Asymptotic?& $\DMLSx_n$ & $\DMLSk{}$ &
			$\Pecper$ & $\Pecpara$ \\ \hline
		(\ref{EQ13_07_27_int}) & No & No & Far & Any & Any & Any \\ \hline
		(\ref{EQ13_05_29}) & No & No & Any & Any & Any & 0 \\ \hline
		(\ref{EQ13_07_27}) & Yes & Yes & Far & Any & Any & Any \\ \hline
		(\ref{EQ13_05_53}) & Yes & Yes & $\neq 0$ & Any & 0 & 0 \\ \hline
		(\ref{EQ13_05_18}) & Yes & No & Far & 0 & 0 & 0 \\ \hline
		(\ref{EQ13_05_38}) & Yes & No & $\neq 0$ & 0 & 0 & 0 \\ \hline
		(\ref{EQ13_05_60}) & Yes & Yes & $\neq 0$ & 0 & 0 & 0 \\ \hline
		(\ref{EQ13_05_54}) & Yes & Yes & $0$ & Any & 0 & 0 \\ \hline
		(\ref{EQ13_05_39}) & Yes & No & $0$ & 0 & 0 & 0 \\ \hline
		\end{tabular}
	}
	\label{table_noise}
	\end{table}
}

\newcommand{\tableParam}[1]{
	\begin{table}[#1]
	\centering
	\caption{System parameters used for numerical and simulation results.}
	{\renewcommand{\arraystretch}{1.4}
		\begin{tabular}{|l|c|c|}
		\hline
		Parameter & Symbol & Value\\ \hline
		Release rate of ideal noise source & $\Ngenavgt{t}$ &
		$1.2\times10^6\, \frac{\molecule}{\second}$ \\ \hline
		Molecules per transmitter emission & $\NemitU{u}$ & $10^4$	\\ \hline
		Probability of transmitter binary $1$ & $\Pone$ 	& $0.5$ 	\\ \hline
		Length of transmitter sequence & $\B{}$	& $100$ bits		\\ \hline
		Transmitter bit interval	& $\TU{u}$		& $0.2\,$ms \\ \hline
		Diffusion coefficient \cite{RefWorks:742,RefWorks:754}
		& $\Dx{\A}$ & $10^{-9}\,\metre^2/\second$ \\ \hline
		Radius of receiver & $\robs$	& $50\,$nm		\\ \hline
		Step size for continuous noise & $\Delta \DMLSt{}$ & $0.1$ \\ \hline
		Step size for transmitters & $\Delta t$ & $2\,\mu\second$ \\ \hline
		\end{tabular}
	}
	\label{table_param}
	\end{table}
}

\newcommand{\tableDMLS}[1]{
	\begin{table}[#1]
	\centering
	\caption{Conversion between dimensional and dimensionless variables. The
	values of $t$, $\vx{}$, and $\kth{}$ correspond to $\DMLSt{}=1$, $\Pec{} = 1$,
	and $\DMLSk{} = 1$, respectively.}
	{\renewcommand{\arraystretch}{1.4}
		\begin{tabular}{|c|c|c|c|c|c|}
		\hline
		$\x_n$ [nm] & $\dist$ [nm] & $t$ [$\mu\second$] &
			$\vx{}$ [$\frac{\metre\metre}{\second}$] &
			$\kth{}$ [$\second^{-1}$] &
			$\NAx{REF}$ \\ \hline
		0 & 50 & 2.5 & 20 & $4\times10^5$ & 3\\ \hline
		50 & 50 & 2.5 & 20 & $4\times10^5$ & 3\\ \hline
		100 & 100 & 10 & 10 & $1\times10^5$ & 12\\ \hline
		200 & 200 & 40 & 5 & $2.5\times10^4$ & 48\\ \hline
		400 & 400 & 160 & 2.5 & $6.25\times10^3$ & 192\\ \hline
		1000 & 1000 & 1000 & 1 & $10^3$ & 1200\\ \hline
		\end{tabular}
	}
	\label{table_dmls}
	\end{table}
}

\newcommand{\figNoiseNoVK}[2]{
	\begin{figure}[#1]
	\centering
	\includegraphics[width=#2\linewidth]
	{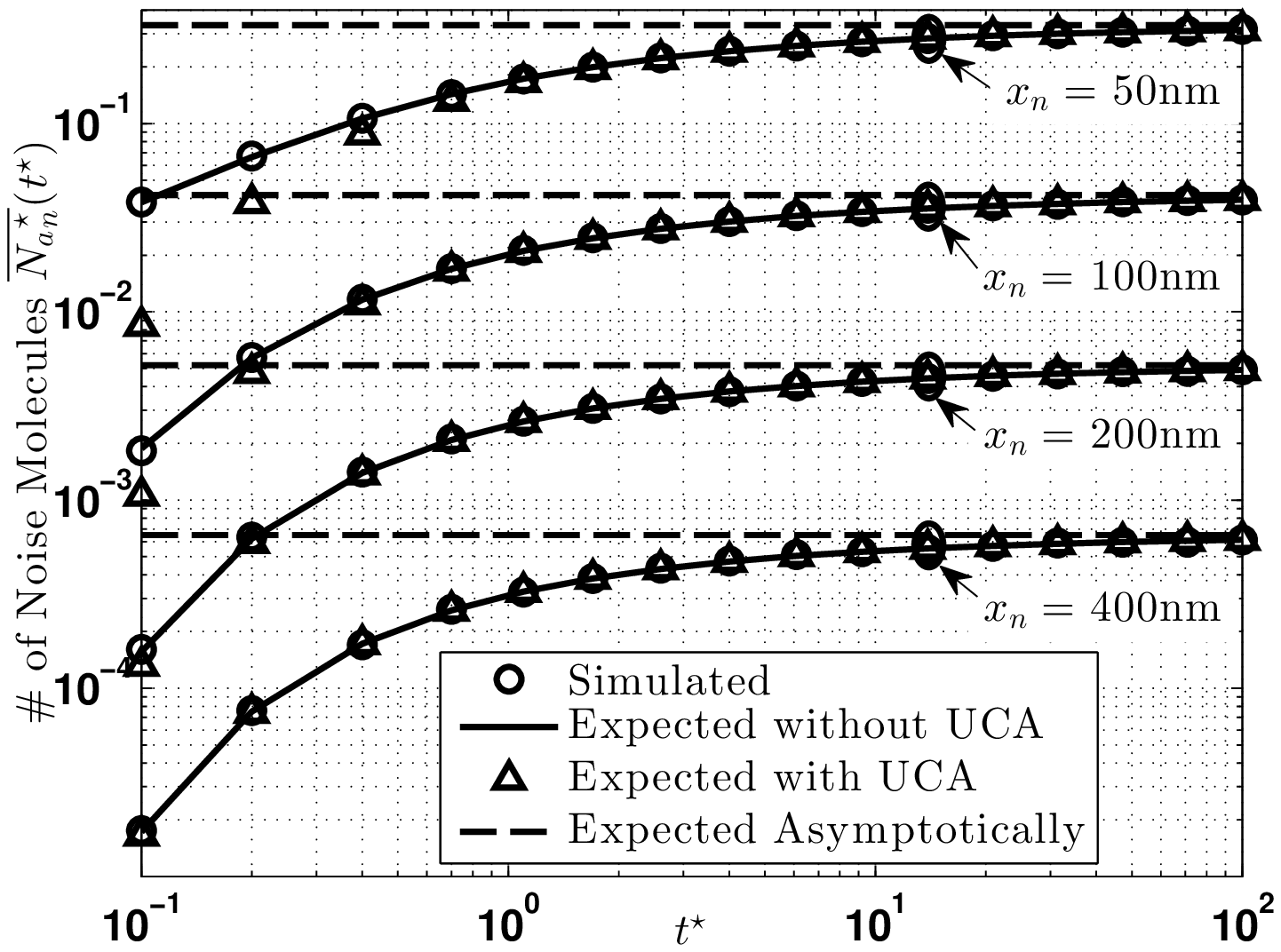}
	\caption{The dimensionless number of noise molecules observed
	at the receiver as a function of time
	when $\Pecpara=\Pecper=0$ and $\DMLSk{}=0$, i.e., when there is no
	advection or molecule degradation.}
	\label{fig_noise_v0_k0}
	\end{figure}
}

\newcommand{\figNoiseKNoV}[2]{
	\begin{figure}[#1]
	\centering
	\includegraphics[width=#2\linewidth]
	{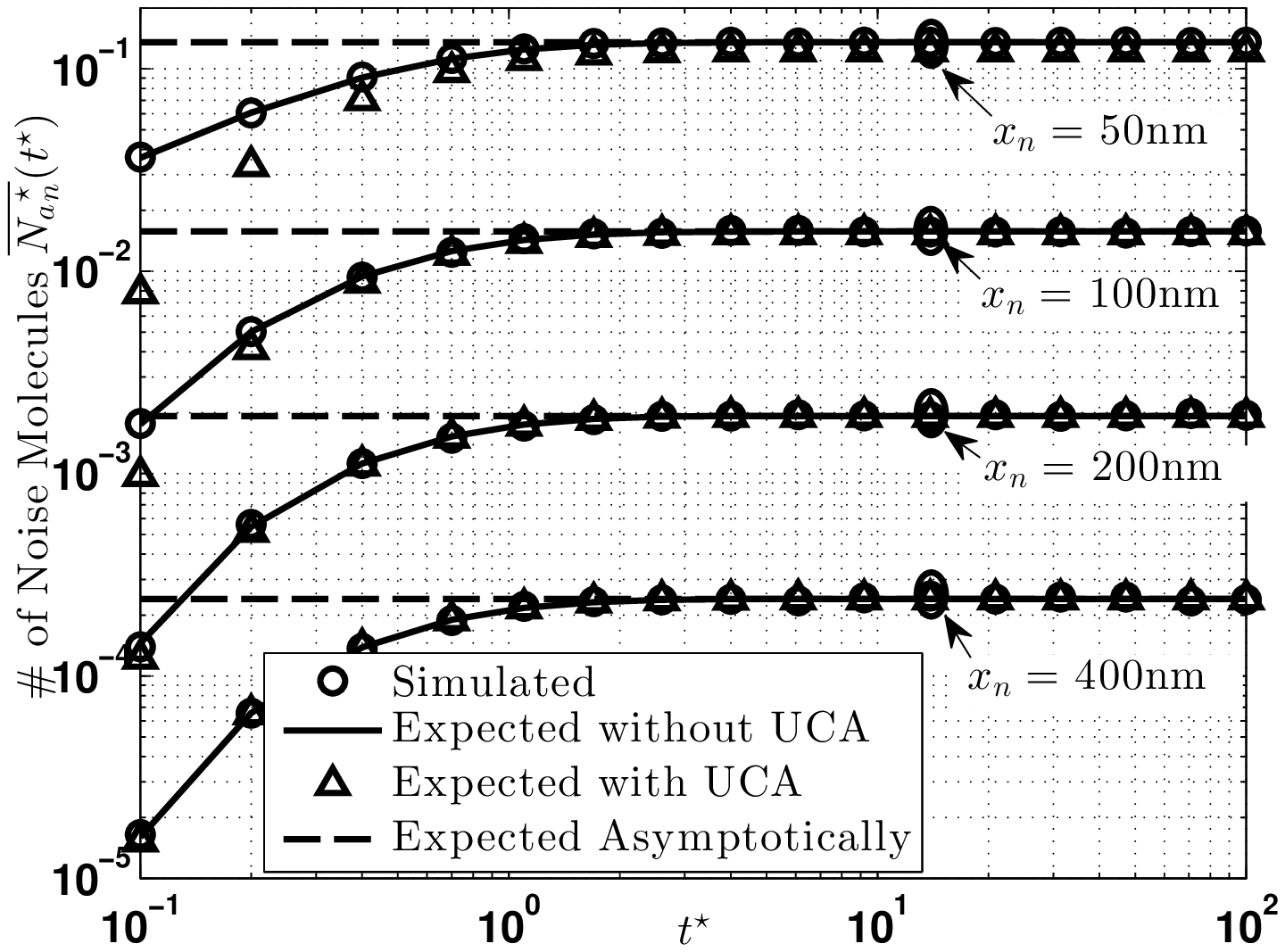}
	\caption{The dimensionless number of noise molecules observed
	at the receiver as a function of time
	when $\Pecpara=\Pecper=0$ and $\DMLSk{} = 1$.}
	\label{fig_noise_v0_k1}
	\end{figure}
}

\newcommand{\figNoiseK}[2]{
	\begin{figure}[#1]
	\centering
	\includegraphics[width=#2\linewidth]
	{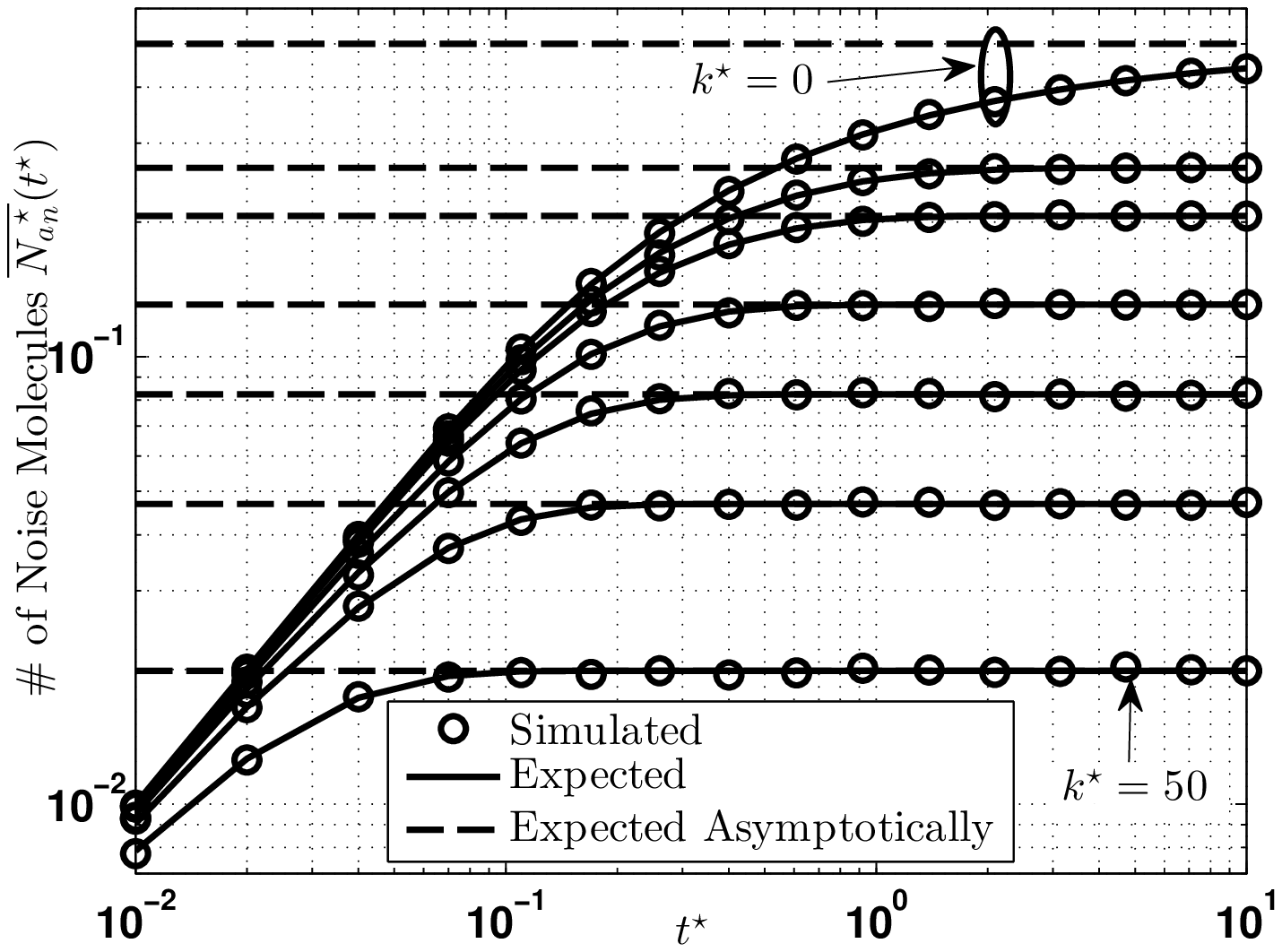}
	\caption{The dimensionless number of noise molecules observed
	at the receiver as a function of time
	when $\x_n = 0$,
	$\Pecpara=\Pecper=0$, and the molecule degradation rate is
	$\DMLSk{} = \{0,1,2,5,10,20,50\}$.}
	\label{fig_noise_k}
	\end{figure}
}

\newcommand{\figNoiseV}[2]{
	\begin{figure}[#1]
	\centering
	\includegraphics[width=#2\linewidth]
	{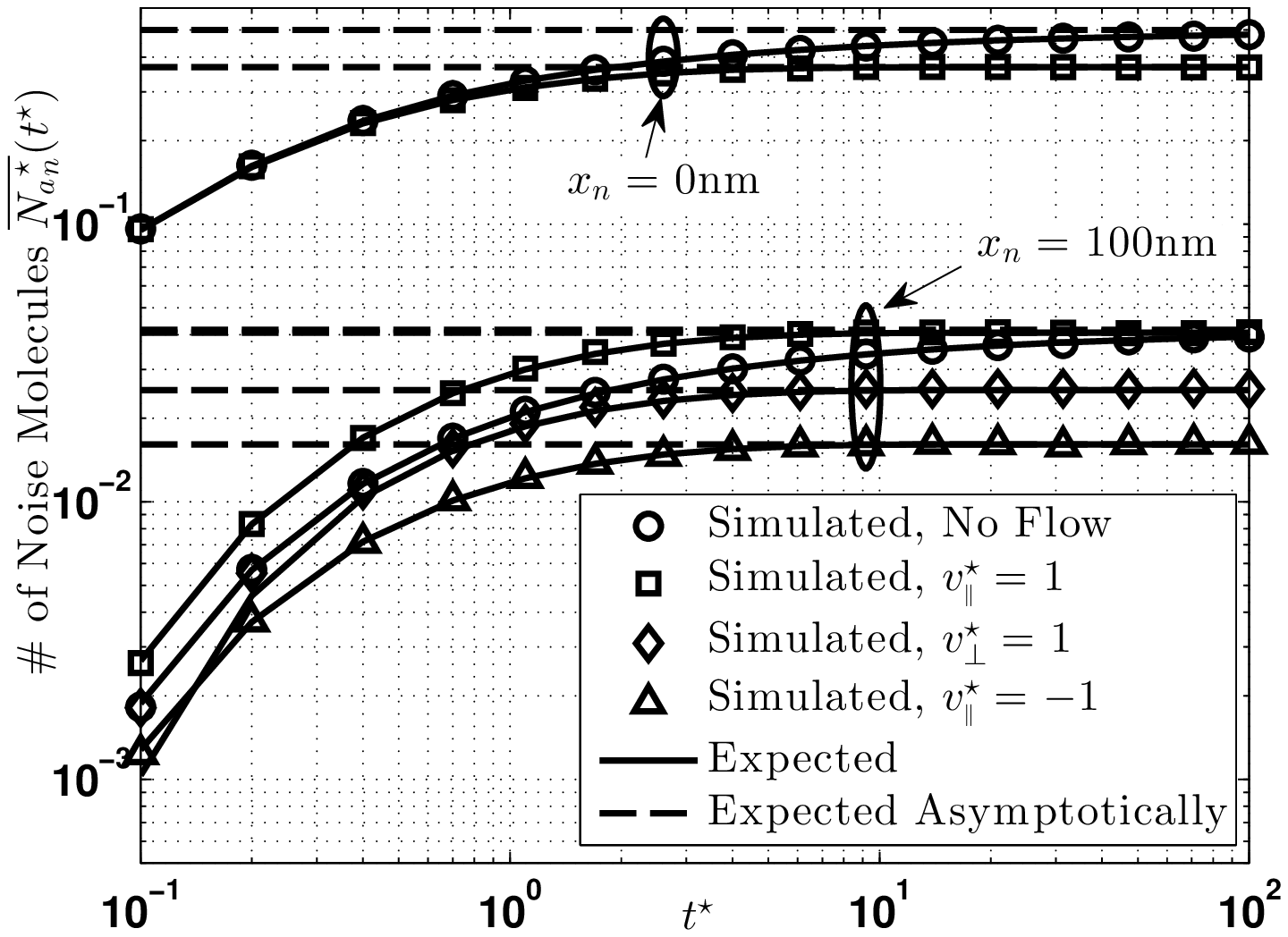}
	\caption{The dimensionless number of noise molecules observed
	at the receiver as a function of time
	when $\DMLSk{}=0$, we vary $\Pecpara$ or $\Pecper$, and we
	consider $\x_n = 0\,$nm and $\x_n = 100\,$nm.}
	\label{fig_noise_v1}
	\end{figure}
}

\newcommand{\figNoiseVFar}[2]{
	\begin{figure}[#1]
	\centering
	\includegraphics[width=#2\linewidth]
	{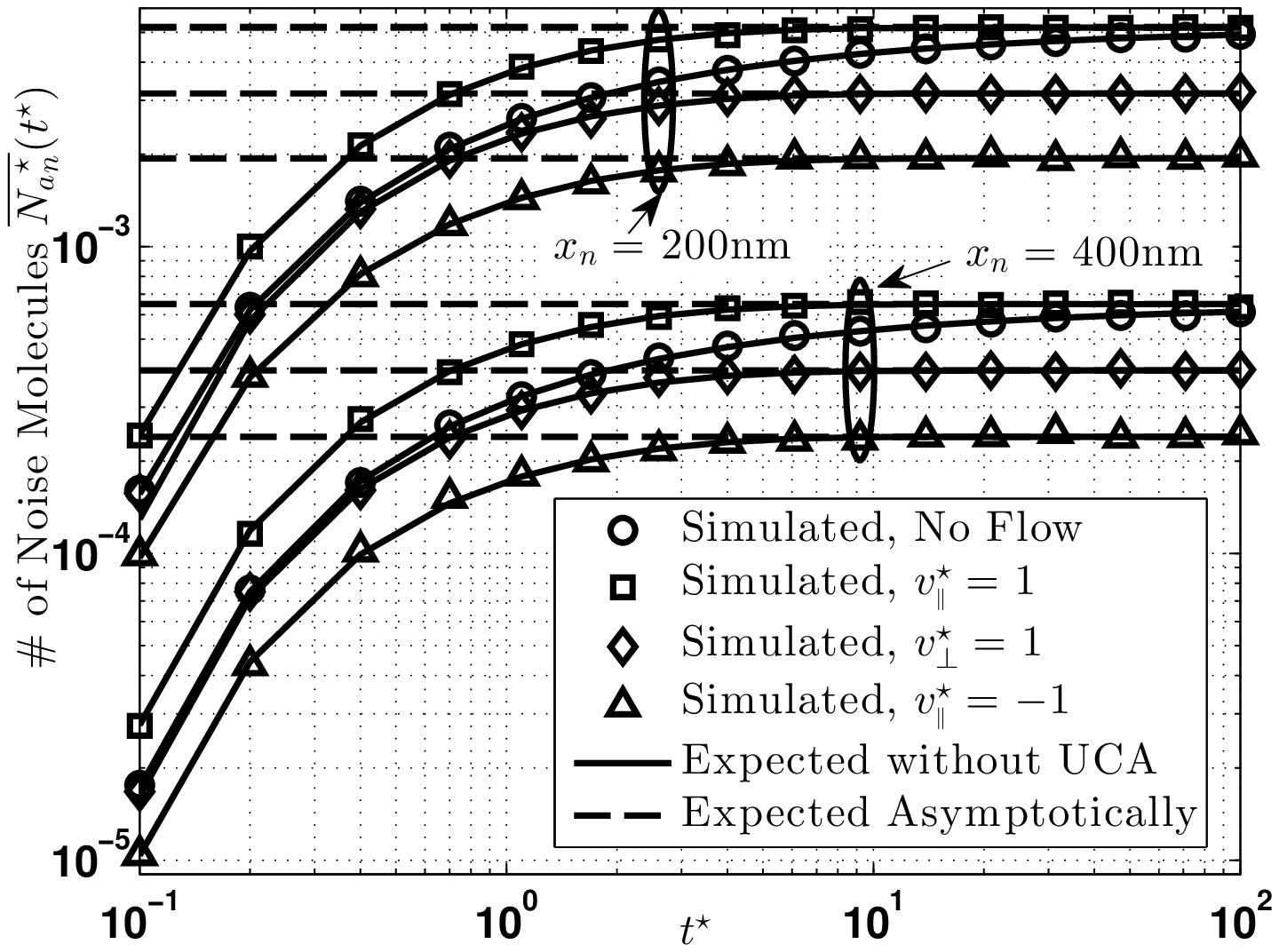}
	\caption{The dimensionless number of noise molecules observed
	at the receiver as a function of time
	when $\DMLSk{}=0$, we vary $\Pecpara$ or $\Pecper$, and we
	consider $\x_n = 200\,$nm and $\x_n = 400\,$nm.}
	\label{fig_noise_v1_far}
	\end{figure}
}

\newcommand{\figInterferer}[2]{
	\begin{figure}[#1]
	\centering
	\includegraphics[width=#2\linewidth]
	{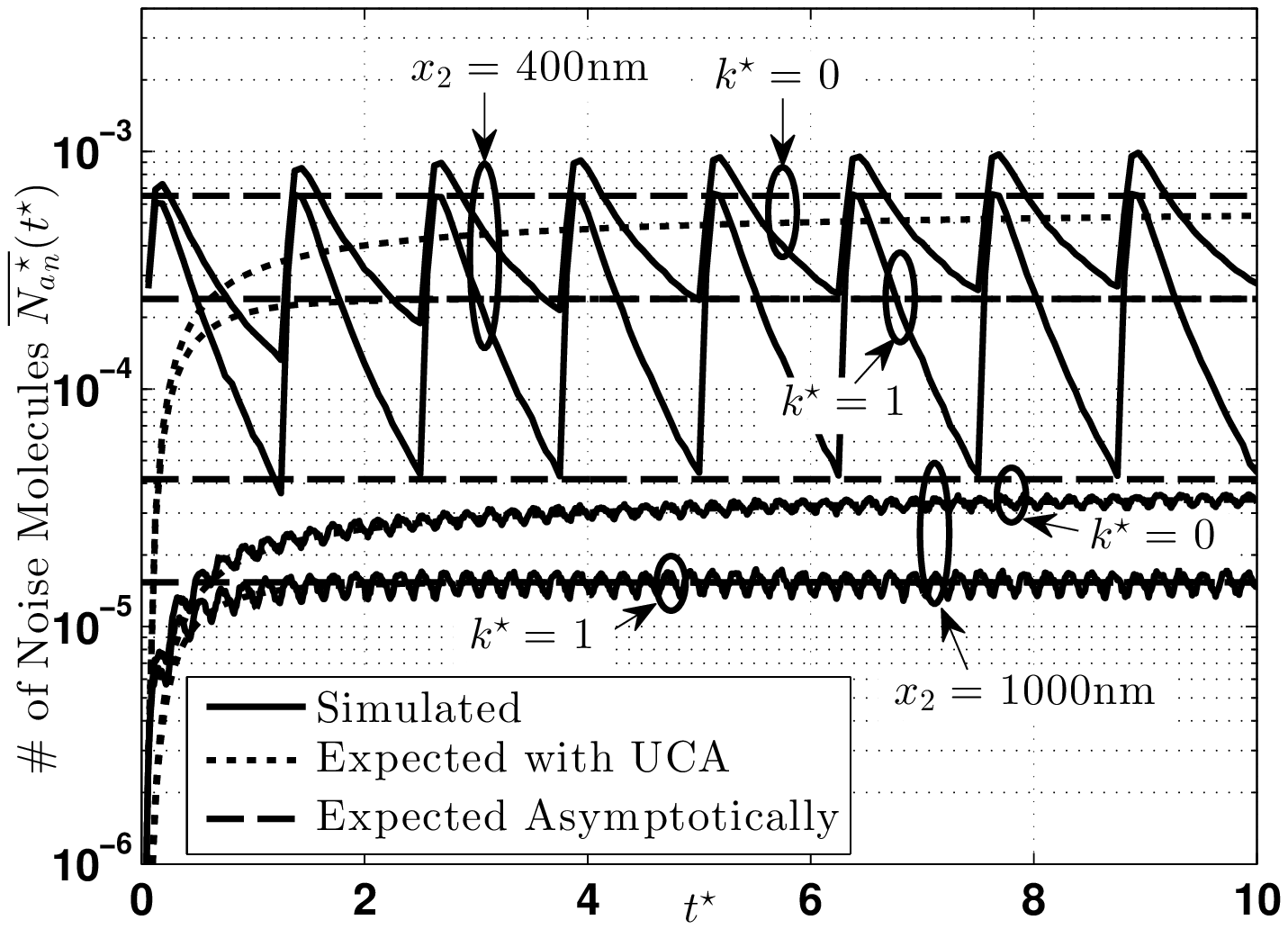}
	\caption{The dimensionless number of interfering molecules observed
	at the receiver as a function of time
	an interfering transmitter placed at
	$\x_2 = 400$nm and $\x_2 = 1\,\mu$m.}
	\label{fig_interferer}
	\end{figure}
}

\newcommand{\figISI}[2]{
	\begin{figure}[#1]
	\centering
	\includegraphics[width=#2\linewidth]
	{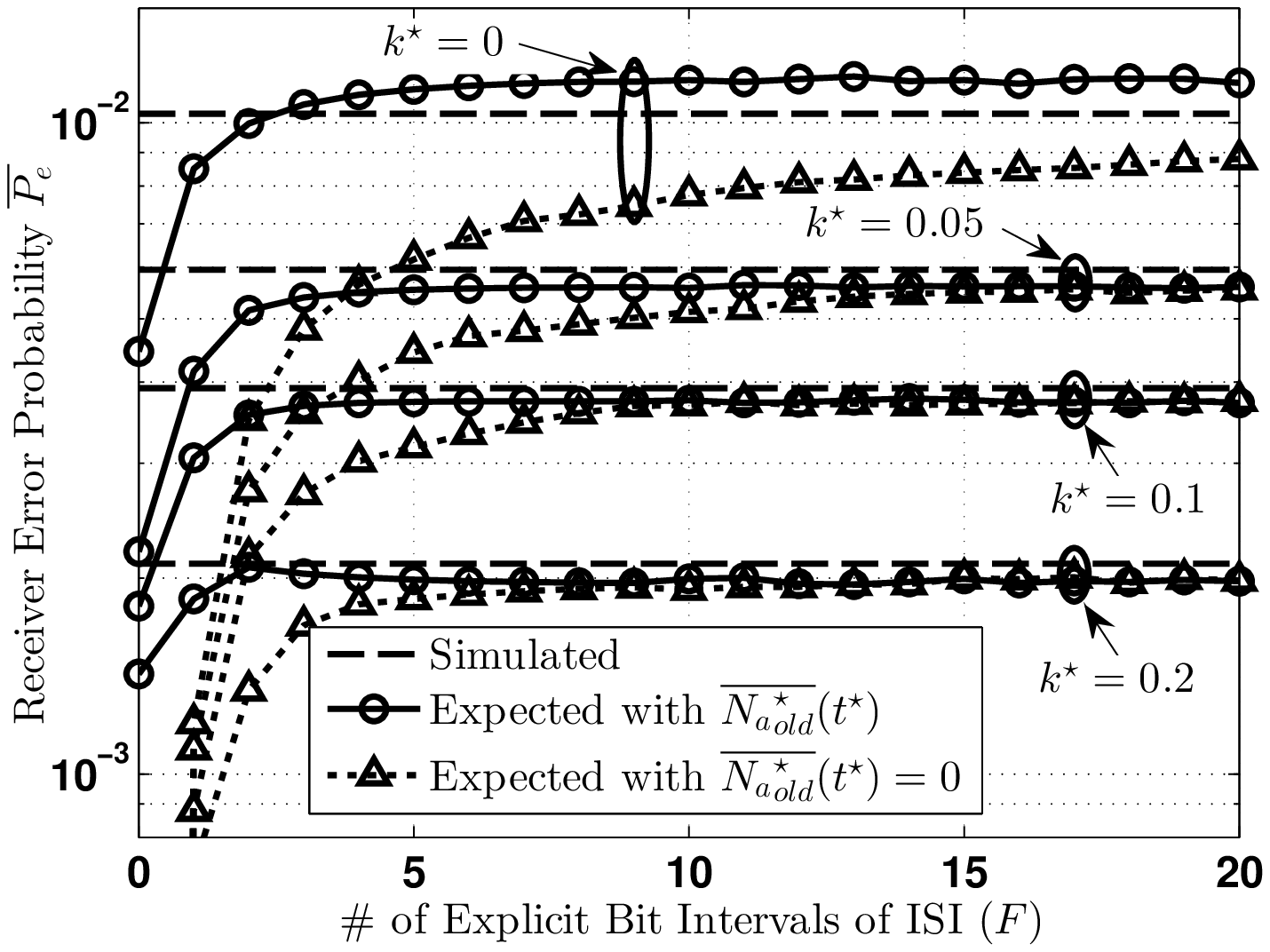}
	\caption{Receiver error probability
	as a function of $\VAmem$,
	the number of bit intervals of \ISI\, treated explicitly,
	for varying
	molecule degradation rate $\DMLSk{}$.}
	\label{fig_isi}
	\end{figure}
}

\maketitle

\begin{abstract}
This paper considers the impact of external noise sources,
including interfering transmitters, on a diffusive
molecular communication system, where the impact is measured
as the number of noise molecules expected to be observed
at a passive receiver. A unifying model for
noise, multiuser interference, and intersymbol interference
is presented, where, under certain circumstances, interference can
be approximated as a noise source that is emitting continuously.
The model includes the presence of advection and molecule
degradation. The time-varying and asymptotic impact
is derived for a series of special cases, some of which facilitate
closed-form solutions. Simulation results show the
accuracy of the expressions derived for the impact of a
continuously-emitting noise source, and show how
approximating old intersymbol interference as a noise source
can simplify the calculation of the expected
bit error probability of a weighted sum detector.
\end{abstract}


\section{Introduction}

\PARstart{M}{olecular} communication is a physical layer
design strategy that could enable the deployment
of nanonetworks by facilitating the sharing of
information between individual devices with nanoscale
functional components. It is envisioned that these
networks will bring new applications to fields that
require diagnostics or actions on a small physical
scale, e.g., healthcare and manufacturing;
see \cite{RefWorks:540, RefWorks:608}. Molecular communication
relies on molecules released by transmitters as information
carriers, and is inspired by the common use of molecules
for information transmission in biological systems;
see \cite{RefWorks:750}.

Passive molecular propagation methods
do not require external energy for transport. The simplest such
method, free diffusion, randomly moves molecules via collisions
with other molecules, and
does not require fixed connections between transceivers. However,
the design of a communication network that is based on free diffusion
faces a number of challenges. The propagation time increases and
the reliability decreases as the distance between transceivers
increases. Intersymbol interference (\ISI) arises if there is no
process to degrade information molecules or carry them away from the
receiver. Furthermore, the deliberate release of molecules by the
intended transmitter might not be the only
local source of information molecules. We refer to other such sources
as \emph{external} molecule sources.
External molecule sources can be
expected in diffusive environments where nanonetworks
may be deployed. Examples include:
\begin{itemize}
    \item Multiuser interference caused by molecules that are emitted by
    the transmitters of other communication links. This interference
    can be mitigated by using
	different molecule types for every communication link, but this might
	not be practical if there is a very large number of links and the
	individual transceivers share a common design.
	\item Unintended leakage from vesicles (i.e., membrane-bound containers)
	where the information molecules are being stored by the transceivers. A
	rupture could result in a steady release of molecules or, if large enough,
	the sudden release of a large number of molecules; see \cite{RefWorks:755}.
	\item The output from an unrelated biochemical process. The biocompatability
	of the nanonetwork may require the selection of a naturally-occurring
	information molecule. Thus, other processes that produce or release that type
	of molecule are effectively noise sources for communication. For example,
	calcium signalling is commonly used as a messenger molecule within cellular
	systems (see \cite[Ch. 16]{RefWorks:588}), so selecting calcium as the
	information carrier in a new molecular communication network deployed in a
	biological environment would mean that the natural occurrence of calcium is a
	source of noise.
	\item The unintended reception of other molecules that are sufficiently
	similar to the information molecules to be recognized by the receiver. For
	example, the receptors at the receiver might not be specific enough to
	only bind to the information molecules,
	or the other molecules might have a shape and size that is very similar
	to that of the information molecules; see \cite[Ch. 4]{RefWorks:588}).
\end{itemize}

Most existing literature on noise analysis in diffusive molecular communication
has considered the noise in the communication link, i.e., via the
noisiness of diffusion itself or chemical
mechanisms at the receiver, cf. e.g. \cite{RefWorks:436, RefWorks:469, RefWorks:644,
RefWorks:687, RefWorks:734, RefWorks:737}, without accounting for the impact of
external noise sources. The impact of multiuser interference
on capacity was evaluated numerically in \cite{RefWorks:656}. Wave theory
was used to approximate both \ISI\, and multiuser interference in
\cite{RefWorks:677}, where \ISI\, was limited to one previous interval and
only multiuser emissions in the current transmission interval were considered.
In \cite{RefWorks:513}, a stochastic model was proposed
that included the spontaneous generation of information molecules in the
propagation environment.

In this paper, we propose a unifying model for external noise sources
(including multiuser interference) and \ISI\, in diffusive
molecular communication. We consider an unbounded physical environment
with steady uniform flow, based on a system model that we studied
in \cite{RefWorks:747, RefWorks:752} (but where we did not develop any
detailed noise analysis; we only assumed that the asymptotic impact
of the noise sources was known). The primary contributions of this paper
are as follows:
\begin{enumerate}
    \item We derive the expected asymptotic (and, wherever possible,
	time-varying) impact of a continuously-emitting noise source, given
	the location of the source and its rate of emission. By impact, we
	refer to the corresponding expected number of molecules observed at the
	receiver, and by asymptotic we refer to the source being active for infinite
	time. Closed-form solutions are available for a number of
	special cases; otherwise, the impact can only be found via numerical integration.
	\item We use asymptotic noise from a source far from the receiver to
	approximate the impact of interfering transmitters, thus providing
	a simple expression for the molecules observed at the receiver due to
	multiuser interference without requiring the interfering transmitters'
	data sequences. The accuracy of this approximation improves as the
	distance between the receiver and the interfering transmitters
	increases.
	\item We approximate ``old'' \ISI\, in the intended
	communication link as asymptotic interference from a
	continuously-emitting source.
	We decompose the received signal into molecules observed due to
	an emission in the current bit interval, molecules that were emitted in
	recent bit intervals, and molecules emitted in older intervals,
	where only the impact of the ``old'' emissions is approximated.
\end{enumerate}

Knowing the expected impact of a noise source enables us to model
its effect on successful transmissions between the intended transmitter
and receiver. For example, in \cite{RefWorks:747, RefWorks:752}
we assumed that we had knowledge of the expected impact of noise
sources in order to evaluate the effect of external
noise on the bit error probability at the intended receiver for a selection
of detectors.
The expected impact of noise sources can also be used to
assess different methods to mitigate the effects of noise, e.g.,
via the degradation of noise molecules as we consider in this paper.

Decomposing the signal received from the intended transmitter
enables us to bridge all existing work on \ISI\, by adjusting the
number of ``recent'' bit intervals and deciding how we analytically model the
``old'' molecules. Most literature on diffusive molecular
communication has accounted for only one recent bit interval and ignored
the impact of old molecules; see
\cite{RefWorks:677,RefWorks:534,RefWorks:574,RefWorks:643}.
More recently, it has become more common to account for \emph{all} molecules
released, i.e., treat all prior bit intervals as recent; see
\cite{RefWorks:667, RefWorks:668, RefWorks:644, RefWorks:687, RefWorks:786} and
our previous work in \cite{RefWorks:662, RefWorks:747}. We introduce the number
of recent bit intervals as a parameter that enables a trade-off between
complexity and accuracy in analyzing receiver performance. Furthermore, modeling
all older \ISI\, as asymptotic noise will be shown to be a more
accurate alternative to assuming that old \ISI\, has no impact at all.

In this paper, we also describe how an asymptotic model for old \ISI\,
simplifies the evaluation of the expected bit error
probability of a weighted sum detector with equal weights.
We proposed this detector as a member of the family of weighted sum detectors
in \cite{RefWorks:747}.
Other possible applications of an asymptotic model for old \ISI\,
include a simplified implementation of the optimal sequence detector (a
detector that we also considered in \cite{RefWorks:747}), or simplifying the
design of an adaptive weight detector, where the decision criteria are adjusted
based on knowledge of the previously received information.

The rest of this paper is organized as follows. The
system model, including the physical environment and its
representation in dimensionless form, is described in Section~\ref{sec_model}.
In Section~\ref{sec_noise}, we derive the time-varying and asymptotic
impact of an external noise source on the receiver.
In Section~\ref{sec_mui}, we
consider the special case of a noise source that is an
interfering transmitter.
We adapt the noise analysis for asymptotic old \ISI\, and use it
to simplify detector performance evaluation in Section~\ref{sec_isi}.
Numerical and simulation results
are described in Section~\ref{sec_num},
and conclusions are drawn in Section~\ref{sec_concl}.

\section{System Model}
\label{sec_model}

We consider an infinite 3-dimensional fluid environment of uniform
constant temperature and viscosity.
The receiver is a sphere with radius $\robs$ and volume $\Vobs$
(if the transmitter is sufficiently far from the receiver, then the precise
shape is irrelevant and we are only interested in $\Vobs$).
As this paper focuses on the impact of unintended sources of information
molecules on the observations made at the receiver, the
receiver is centered at the origin. Without loss of generality,
the intended transmitter is placed at coordinates $\{-\x_1,0,0\}$.
We assume there is steady uniform flow (or drift) in an arbitrary direction
with a velocity component in each dimension, i.e.,
$\vxvec{} = \{\vx{\x},\vx{\y},\vx{\z}\}$.

The receiver is a passive observer that does not impede diffusion
or initiate chemical reaction (so that we can focus on the impact
of the propagation environment). Its only interaction with the environment
is the perfect counting of $\A$ molecules if they are within
$\Vobs$; any other molecules that might be present are ignored.
$\A$ molecules are the information molecules that can be emitted by the
transmitter or by some other sources. In practice, the receiver
would observe $\A$ molecules by having them bind to receptors that
are on the surface of or inside $\Vobs$.

The expected local concentration of $\A$ molecules at the point defined by
vector $\rad{}$ and at time $t$ in
$\molecule\cdot\metre^{-3}$ is $\CxFun{\A}{\rad{}}{t}$, and we write
$\Cx{\A}$ for compactness.
All $\A$ molecules diffuse
independently with constant diffusion coefficient $\Dx{\A}$,
and they can degrade into a form that cannot be detected by the
receiver via a reaction mechanism that can be described as
\begin{equation}
\label{k1_mechanism}
\A \xrightarrow{\kth{}} \emptyset,
\end{equation}
where $\kth{}$ is the reaction rate constant in $\second^{-1}$.
If $\kth{} = 0$, then this degradation is negligible.
Eq. (\ref{k1_mechanism}) is a first-order reaction, but it can also
be used to approximate higher-order reactions or reaction mechanisms
with multiple steps. For example, in our previous work where we considered
enzymes in the propagation environment to mitigate \ISI, we
implicitly used (\ref{k1_mechanism}) to derive a bound on the
expected number of observed molecules;
see \cite{RefWorks:631, RefWorks:662}. First-order reactions
have also been used to approximate higher-order reactions in a molecular
communication context in \cite{RefWorks:737}, where the reactions
occurred only at the receiver.

We emphasize that our assumptions include
a passive receiver, first-order $\A$ molecule
degradation throughout the environment, and the constant diffusion of $\A$
molecules.
These assumptions make our model analytically tractable but ignore the impact of
effects including anomalous diffusion, localized chemical reactions, and other
interactions between molecules. We are interested in studying such complex
systems in our future work.

For clarity of exposition in the remainder of this paper,
we convert our system model into dimensionless form.
We have used dimensional analysis in our previous work, including
\cite{RefWorks:752,RefWorks:706}, because it generalizes our model's
scalability and facilitates comparisons between different dimensional
parameter sets. In this paper, dimensional analysis also provides
clarity of exposition by reducing the number of parameters that
appear in the equations.
Unless otherwise noted, all variables that are described in this paper
are assumed to be dimensionless
(as denoted by a ``$\star$'' superscript), and they
are equal to the dimensional variables scaled by the
appropriate reference variables; see \cite{RefWorks:633}
for more on dimensional analysis.

We define reference distance $\dist$ in $\metre$ and reference number
of molecules $\NAx{REF}$. We also define
reference concentration $\Cx{0} = \NAx{REF}/\dist^3$ in
$\molecule\cdot\metre^{-3}$.
We then define the dimensionless concentration of $\A$ molecules
as $\DMLSC{\DMLSA} = \Cx{\A}/\Cx{0}$, dimensionless time
as $\DMLSt{} = \Dx{\A}t/\dist^2$, and the dimensionless reaction rate
constant as $\DMLSk{} = \dist^2\kth{}/\Dx{\A}$. The
dimensionless coordinates along the three axes are
\begin{equation}
\label{AUG12_43_coor}
\DMLSx = \frac{\x}{\dist}, \quad
\DMLSy = \frac{\y}{\dist}, \quad
\DMLSz = \frac{\z}{\dist},
\end{equation}
such that they are the dimensional coordinates scaled (i.e.,
normalized) by the reference distance $\dist$.
Advection is
represented dimensionlessly with the Peclet number, $\Pec{}$, written as
\cite[Ch. 1]{RefWorks:750}
\begin{equation}
\label{EQ13_07_30}
\Pec{} = \frac{\vx{}\dist}{\Dx{\A}},
\end{equation}
where $\vx{} = |\vxvec{}|$ is the speed of the fluid. $\Pec{}$ measures the
relative impact of advection versus diffusion on molecular transport. If
$\Pec{} = 1$, then the typical time for a molecule to diffuse the reference
distance $\dist$, i.e., $\dist^2/\Dx{\A}$, is equal
to the typical time for a molecule to move the same distance by advection alone.
A value of $\Pec{}$ much less or much greater than $1$ signals the dominance
of diffusion or advection, respectively. We saw in \cite{RefWorks:752} that
the impact of steady uniform flow on successful communication
also depends on the \emph{direction} of flow, so
we define $\Pec{}$ along each
dimension as
\begin{equation}
\label{EQ13_07_29}
\Pecpara = \frac{\vx{\x}\dist}{\Dx{\A}}, \quad
\Pec{\perp,1} = \frac{\vx{\y}\dist}{\Dx{\A}}, \quad
\Pec{\perp,2} = \frac{\vx{\z}\dist}{\Dx{\A}}.
\end{equation}

The dimensionless signal observed at the receiver, $\DMLSNxt{\DMLSt{}}{obs}$,
is the cumulative impact of all
molecule emitters in the environment, including interfering transmitters
and other noise sources.
Due to the independence of the diffusion of all $\A$ molecules, we can
apply superposition to the
impacts of the individual sources, such that the cumulative impact of
multiple noise sources is the sum of the impacts
of the individual sources.
If we assume that there are $\U-1$ sources of
$\A$ molecules that are not the intended transmitter (without specifying
what kinds of sources these are, i.e., other transmitters or
simply ``leaking'' $\A$ molecules), then the complete observed signal
can be written as
\begin{equation}
\DMLSNxt{\DMLSt{}}{obs} = \DMLSNxt{\DMLSt{}}{1} +
\sum_{u=2}^{\U}\DMLSNxt{\DMLSt{}}{u},
\label{EQ13_05_28_obs}
\end{equation}
where $\DMLSNxt{\DMLSt{}}{1}$ is the signal from the intended transmitter.
Without loss of generality (because the impact of multiple
molecule sources can be superimposed), we can analyze each
$\DMLSNxt{\DMLSt{}}{u}$ term independently. Thus, for
clarity, we assume that the $u$th source is placed at
$\{-\DMLSx_u,0,0\}$, where $\DMLSx_u \ge 0$.
Furthermore,
for all molecule
sources in (\ref{EQ13_05_28_obs}), the advection variables
must be defined relative to the source's corresponding coordinate frame, such
that $\Pecpara>0$ always represents flow from the source towards the receiver.
By symmetry, and without loss of generality, we can set $\Pec{\perp,2} = 0$ and
write $\Pec{\perp,1} = \Pecper$, such that $\Pecper$ represents
flow perpendicular to the line between the source and receiver.

In Table~\ref{table_overview}, we summarize where the different terms in
(\ref{EQ13_05_28_obs}) are analyzed in the remainder of this paper and whether
each type of source is treated as continuously-emitting.
In Sections \ref{sec_noise} and \ref{sec_mui},
we model $\DMLSNxt{\DMLSt{}}{u}$ as a random noise source and as an interfering
transmitter, respectively. In Section \ref{sec_isi}, we decompose
$\DMLSNxt{\DMLSt{}}{1}$ to approximate old \ISI\, as asymptotic
noise.

\ifOneCol
\else
	\tableOverview{!tb}{1.65cm}{1.35cm}{0.8cm}{2cm}
\fi

\section{External Additive Noise}
\label{sec_noise}

In this section, we derive the impact of external noise sources
on the receiver, given that we have some knowledge about the
location of the noise sources and their mode of emission.
First, we consider a single point noise source placed at
$\{-\DMLSx_n,0,0\}$ where $\DMLSx_n$ is non-negative (we change the
subscript of the source from $u$ to $n$ in order to emphasize that
this source is random noise and not a transmitter of information).
The source emits molecules
according to the random process $\Ngent{t}$, represented dimensionlessly
as $\DMLSNgent{\DMLSt{}} = \dist^2\Ngent{t}/\left(\Dx{\A}\NAx{REF}\right)$.
Assuming that
the expected generation of molecules can be represented as a step function, i.e.,
$\Ngenavgt{t} = \Ngen, t \ge 0$, we then formulate the expected impact
of the noise source at the receiver, $\DMLSNntavg{\DMLSt{}}$.
In its most general form,
we will not have a closed-form solution for the expected impact of
the noise source. Next, we present
either time-varying or asymptotic expressions
for a number of relevant special cases, some of which are in closed form
and others that facilitate numerical integration. While we are
ultimately most interested in asymptotic solutions (particularly
for extension to the analysis of interference), time-varying solutions
are also of interest
when they are available because they give us insight into how long
a noise source must be ``active'' before we can model its impact as
asymptotic. Time-varying solutions will also be useful when
we consider old \ISI\,
in Section~\ref{sec_isi}.
As previously noted, we can use superposition to consider the cumulative
impact of multiple noise sources, as given in (\ref{EQ13_05_28_obs}),
where the advection variables $\Pecpara$ and $\Pecper$ must be
defined for each source
depending on its location.

\subsection{General Noise Model}

First, we require the channel impulse response due to the noise source,
i.e., the \emph{expected} concentration of molecules observed at the
receiver due to an emission of one molecule by the noise source at $\DMLSt{} =
0$. This is analogous
to the channel impulse response due to an intended transmitter at
the same location. The reaction-diffusion differential
equation describing the expected motion of $\A$ molecules can be
written by applying the principles of chemical kinetics
(see \cite[Ch. 9]{RefWorks:585}) to (\ref{k1_mechanism}) and
including the advection terms (as in \cite[Ch. 4]{RefWorks:630}), i.e.,
\begin{equation}
\label{JUN12_33_DMLS}
\pbypx{\DMLSC{\DMLSA}}{\DMLSt{}} = \nabla^2\DMLSC{\DMLSA}
- \Pecpara\pbypx{\DMLSC{\DMLSA}}{\DMLSx}
- \Pecper\pbypx{\DMLSC{\DMLSA}}{\DMLSy}
- \DMLSk{}\DMLSC{\DMLSA},
\end{equation}
where
\ifOneCol
\begin{equation}
\pbypx{\DMLSC{\DMLSA}}{\DMLSt{}} = \pbypx{\Cx{\A}}{t}\frac{\dist^2}{\Dx{\A}\Cx{0}},
\quad \nabla^2\DMLSC{\DMLSA} = \frac{\dist^2}{\Cx{0}}\nabla^2\Cx{\A},
\quad \pbypx{\DMLSC{\DMLSA}}{\DMLSx} =
\pbypx{\Cx{\A}}{\x}\frac{\dist}{\Cx{0}},
\quad \pbypx{\DMLSC{\DMLSA}}{\DMLSy} = \pbypx{\Cx{\A}}{\y}\frac{\dist}{\Cx{0}},
\label{AUG12_45_v}
\end{equation}
\else
\begin{align}
\pbypx{\DMLSC{\DMLSA}}{\DMLSt{}} =
&\;\pbypx{\Cx{\A}}{t}\frac{\dist^2}{\Dx{\A}\Cx{0}}, \quad
\nabla^2\DMLSC{\DMLSA} = \frac{\dist^2}{\Cx{0}}\nabla^2\Cx{\A},
\nonumber \\
\label{AUG12_45_v}
\pbypx{\DMLSC{\DMLSA}}{\DMLSx} = &\;
\pbypx{\Cx{\A}}{\x}\frac{\dist}{\Cx{0}}, \quad
\pbypx{\DMLSC{\DMLSA}}{\DMLSy} = \pbypx{\Cx{\A}}{\y}\frac{\dist}{\Cx{0}},
\end{align}
\fi
and it is straightforward (using a moving reference frame) to
show that the channel impulse response at the point
$\{\DMLSx,\DMLSy,\DMLSz\}$ due to the noise source at
$\{-\DMLSx_n,0,0\}$ is
\begin{equation}
\label{APR12_22}
\DMLSC{\DMLSA} = \frac{1}{(4\pi
\DMLSt{})^{3/2}}\EXP{\frac{-\DMLSradmag{}^2}{4 \DMLSt{}}
-\DMLSk{}\DMLSt{}},
\end{equation}
where
$\DMLSradmag{}^2 = (\DMLSx + \DMLSx_n - \Pecpara\DMLSt{})^2 +
(\DMLSy - \Pecper\DMLSt{})^2 + (\DMLSz)^2$
is the square of the time-varying \emph{effective} distance between the
noise source and the point $\{\DMLSx,\DMLSy,\DMLSz\}$.

Unlike an intended transmitter, the noise source is emitting
molecules as described by the general random process $\DMLSNgent{\DMLSt{}}$.
We are already averaging over the randomness of the diffusion channel
(i.e., we have the expected channel impulse response), so we
only consider the time-varying mean of the noise source
emission process, i.e., $\DMLSNgenavgt{\DMLSt{}}$.
Thus, the expected impact of the noise source is found by multiplying
(\ref{APR12_22}) by $\DMLSNgenavgt{\DMLSt{}}$, integrating over
$\DMLSV$, and then integrating over all time up to $\DMLSt{}$, i.e.,
\begin{equation}
\label{APR12_42_DMLS}
\DMLSNntavg{\DMLSt{}} = \int\limits_{-\infty}^{\DMLSt{}}\!
\int\limits_0^{\DMLSr{obs}}
\int\limits_{0}^{2\pi}
\int\limits_{0}^{\pi}
{\DMLSr{i}}^2\DMLSNgenavgt{\DMLStau}
\DMLSC{\DMLSA}\sin\theta
d\theta d\phi d\DMLSr{i}d\DMLStau,
\end{equation}
where $\DMLSr{i}$ is the magnitude of the distance from
the origin to the point $\{\DMLSx,\DMLSy,\DMLSz\}$ within $\DMLSV$.
To solve (\ref{APR12_42_DMLS}),
we must also convert $\DMLSradmag{}^2$ from cartesian to spherical
coordinates, which can be shown to be
\ifOneCol
\begin{align}
\DMLSradmag{}^2 = &\;
{\DMLSr{i}}^2 + {\DMLSx_0}^2 - 2\DMLSt{}\DMLSx_0\Pecpara
+ 2\DMLSx_0\DMLSr{i}\cos\phi\sin\theta
+ {\DMLSt{}}^2\left({\Pecpara}^2
+ {\Pecper}^2\right) \nonumber \\
& -2\DMLSt{}\DMLSr{i}
\left(\Pecpara\cos\phi\sin\theta + \Pecper\sin\phi\sin\theta\right),
\label{EQ13_07_25}
\end{align}
\else
\begin{align}
\DMLSradmag{}^2 = &\,
{\DMLSr{i}}^2 + {\DMLSx_0}^2 - 2\DMLSt{}\DMLSx_0\Pecpara
+ 2\DMLSx_0\DMLSr{i}\cos\phi\sin\theta
\nonumber \\
& -2\DMLSt{}\DMLSr{i}
\left(\Pecpara\cos\phi\sin\theta + \Pecper\sin\phi\sin\theta\right)
\nonumber \\
&  + {\DMLSt{}}^2\left({\Pecpara}^2
+ {\Pecper}^2\right),
\label{EQ13_07_25}
\end{align}
\fi
where $\phi = \tan^{-1}\left(\DMLSy/\DMLSx\right)$ and $\theta =
\cos^{-1}\left(\DMLSz/\DMLSr{i}\right)$.
Generally, (\ref{APR12_42_DMLS}) does not have a known closed-form
solution, even if we omit the integral over time.
In the following subsection, we present a series of cases for which
(\ref{APR12_42_DMLS}) can be more easily solved (numerically or in
closed form),
for either arbitrary $\DMLSt{}$
or as $\DMLSt{} \to \infty$. In the asymptotic case, we write
$\DMLSNntavg{\DMLSt{}}\big|_{\DMLSt{} \to \infty} = \DMLSNnavg$
for compactness. The asymptotic case will also be useful to
approximate multiuser
interference and old intersymbol interference in
Sections \ref{sec_mui} and \ref{sec_isi}, respectively.

\subsection{Tractable Noise Analysis}

For tractability, we will assume throughout the remainder of this section
that the \emph{expected} noise source emission in (\ref{APR12_42_DMLS})
can be described as a step function, i.e.,
$\DMLSNgenavgt{\DMLSt{}} = \dist^2\Ngen/\left(\Dx{\A}\NAx{REF}\right),
\DMLSt{} \ge 0$.
Furthermore, we choose $\NAx{REF} = \dist^2\Ngen/\Dx{\A}$, so that
$\DMLSNgenavgt{\DMLSt{}} = 1, \DMLSt{} \ge 0$
(the case where the noise source also ``shuts off'' at some future time,
for example when a ruptured vesicle is depleted, is an interesting
one that we leave for future work). We note that the emission of molecules
by the noise source could then be deterministically uniform, such
that the emission process $\DMLSNgent{\DMLSt{}}$ is in
fact $\DMLSNgenavgt{\DMLSt{}}$ (e.g.,
via leakage from a vesicle that ruptured at $\DMLSt{} = 0$),
or it could be random with independent emission times (e.g., the
stochastic output of a chemical reaction mechanism
with a constant expected generation rate
that was triggered to begin at $\DMLSt{} = 0$). Strictly speaking, in
the latter case the \emph{expected} emission rate is $1$.
This will not affect any of the following analysis because we
are deriving the \emph{expected} impact $\DMLSNntavg{\DMLSt{}}$.
We emphasize that our analysis focuses on the expected impact and not
the complete probability density function (\PDF) of the impact.
A case-by-case analysis of the noise release statistics
would be needed to determine the time-varying \PDF\, of the
impact at the receiver.

The solutions to (\ref{APR12_42_DMLS}) that we present in the remainder
of this section follow one of two general strategies. Both strategies reduce
(\ref{APR12_42_DMLS}) to a single integral, which can be solved numerically or
reduced to closed form if additional assumptions are made.
The first strategy is the uniform
concentration assumption (\UCA), where we assume that the expected
concentration of $\A$ molecules due to the noise source is uniform and equal to
that expected at the center of the receiver (i.e., at the origin). This
assumption is accurate if the noise source is sufficiently far from the
receiver, such that the expected concentration of $\A$ molecules will not vary
significantly throughout the receiver. We studied the accuracy of this
assumption for a transmitter using impulsive binary-coded modulation in the
presence of steady uniform flow in \cite{RefWorks:752}. Here,
applying the \UCA\, means that we do not need to
integrate over $\DMLSV$ and (\ref{APR12_42_DMLS}) becomes
\begin{equation}
\label{EQ13_07_27_int}
\DMLSNntavg{\DMLSt{}} = 
\DMLSV\int\limits_{0}^{\DMLSt{}}
\DMLSCxFun{\DMLSA}{\DMLSr{eff}}{\DMLStau}
d\DMLStau,
\end{equation}
where ${\DMLSr{eff}}^2 = (\DMLSx_n - \Pecpara\DMLStau)^2 +
(\Pecper\DMLStau)^2$ is the square
of the \emph{effective} distance from the noise source to the receiver, and
the expected concentration at the receiver is
\begin{equation}
\label{EQ13_07_26}
\DMLSCxFun{\DMLSA}{\DMLSr{eff}}{\DMLSt{}} = 
\frac{1}{(4\pi\DMLSt{})^{3/2}}
\EXP{-\frac{{\DMLSr{eff}}^2}
{4 \DMLSt{}} - \DMLSk{}\DMLSt{}}.
\end{equation}

The second strategy for solving (\ref{APR12_42_DMLS}) does not apply the
\UCA, so we include the integration over $\DMLSV$.
We considered that integration (but without the integration over time)
for no molecule degradation
in \cite{RefWorks:752}, and for general advection
we could only evaluate the integral over $\DMLSr{i}$.
However, a closed-form solution was possible if $\Pecper = 0$, which
we derived in \cite{RefWorks:752} from the no-flow case presented in
\cite[Th. 2]{RefWorks:706} via a change of variables. Including the
integration over time and the impact of molecule degradation,
where $\DMLSNgent{\DMLSt{}} = 1, \DMLSt{} \ge 0$,
\cite[Eq. (17)]{RefWorks:752} becomes
\ifOneCol
\begin{align}
\DMLSNntavg{\DMLSt{}} = &\; \int\limits_0^{\DMLSt{}} \Bigg\{
\frac{1}{2}\left[\ERF{\frac{\DMLSr{obs}\!-
\DMLSrad{eff}}{2{\DMLStau}^\frac{1}{2}}} +
\ERF{\frac{\DMLSr{obs}\!+\DMLSrad{eff}}{2{\DMLStau}^\frac{1}{2}}}\right] \nonumber \\
& + \frac{1}{\DMLSrad{eff}}\sqrt{\frac{\DMLStau}{\pi}}
\Bigg[\EXP{-\frac{(\DMLSrad{eff}+\DMLSr{obs})^2}{4\DMLStau}}
- \EXP{-\frac{(\DMLSrad{eff}-\DMLSr{obs})^2}{4\DMLStau}}\Bigg]
\Bigg\}\EXP{- \DMLSk{}\DMLStau}d\DMLStau,
\label{EQ13_05_29}
\end{align}
\else
\begin{align}
\DMLSNntavg{\DMLSt{}} = &\; \int\limits_0^{\DMLSt{}} \Bigg\{
\frac{1}{2}\left[\ERF{\frac{\DMLSr{obs}\!-
\DMLSrad{eff}}{2{\DMLStau}^\frac{1}{2}}} \!+
\ERF{\frac{\DMLSr{obs}\!+\DMLSrad{eff}}{2{\DMLStau}^\frac{1}{2}}}\!\right] \nonumber \\
& + \frac{1}{\DMLSrad{eff}}\sqrt{\frac{\DMLStau}{\pi}}
\Bigg[ \EXP{-\frac{(\DMLSrad{eff}+\DMLSr{obs})^2}{4\DMLStau}}\nonumber \\
& - \EXP{-\frac{(\DMLSrad{eff}-\DMLSr{obs})^2}{4\DMLStau}}\Bigg]
\Bigg\}\EXP{- \DMLSk{}\DMLStau}d\DMLStau,
\label{EQ13_05_29}
\end{align}
\fi
where $\DMLSrad{eff} = -\left(\DMLSx_n - \Pecpara\DMLStau\right)$ is
the effective distance along the $\DMLSx$-axis
from the noise source to the center of the receiver, and
the error function is \cite[Eq. 8.250.1]{RefWorks:402}
\begin{equation}
\label{APR12_32}
\ERF{a} = \frac{2}{\sqrt{\pi}}\int_0^a \EXP{-b^2} db.
\end{equation}

Eq. (\ref{EQ13_05_29})
can be evaluated numerically but, unlike (\ref{EQ13_07_27_int}), is valid
for \emph{any} $\DMLSx_n$ (although special consideration must be made if
$\DMLSx_n = 0$, i.e., the ``worst-case'' location
for the noise source, and we consider that case at the end of this
subsection).

The two strategies that we have presented reduce (\ref{APR12_42_DMLS}) to
a single integral (either (\ref{EQ13_07_27_int}) or (\ref{EQ13_05_29})), thereby
facilitating numerical evaluation. In the remainder of this subsection, we make
additional assumptions that enable us to solve (\ref{APR12_42_DMLS}) in closed
form.

\subsubsection{Asymptotic Solutions}

In the asymptotic case, i.e., as $\DMLSt{} \to \infty$, it is
straightforward to show that (\ref{EQ13_07_27_int}) becomes
\begin{equation}
\DMLSNnavg
= \frac{\DMLSV}{4\pi\DMLSx_n}
\EXP{\frac{\DMLSx_n\Pecpara}{2}
- \frac{\DMLSx_n}{2}\sqrt{{\Pecpara}^2 + {\Pecper}^2
+4\DMLSk{}}},
\label{EQ13_07_27}
\end{equation}
where we apply \cite[Eq. 3.472.5]{RefWorks:402}
\begin{equation}
\int\limits_{0}^{\infty} \frac{1}{a^{3/2}}\EXP{-ba-\frac{c}{a}}da =
\sqrt{\frac{\pi}{c}}\EXP{-2\sqrt{bc}},
\label{EQ13_05_20}
\end{equation}
and recall that $\DMLSx_n$ is positive.

\begin{remark}
\label{remark_far_flow}
From (\ref{EQ13_07_27}) it can be shown that, if there is
no flow in the $\y$-direction and no molecule degradation (i.e.,
$\Pecper = 0$ and $\DMLSk{} = 0$), then any \emph{positive} flow along
the $\x$-direction (i.e., $\Pecpara > 0$) will not change the
asymptotic impact of the noise source. We had expected that this
flow would increase the asymptotic impact in comparison to the no-flow case,
so this is a somewhat surprising result.
\end{remark}

An asymptotic closed-form solution to (\ref{EQ13_05_29}) is possible if we impose
$\Pecpara = 0$, such that we are restricted to the no-flow case. If the noise source
is also close to the receiver, then this is another ``worst-case''
scenario because there is no advection to carry the noise molecules away. The
result is presented in the following theorem:

\begin{theorem}[$\DMLSNnavg$ in Absence of Flow]
\label{theorem_no_flow}
The expected asymptotic impact (i.e., as $\DMLSt{} \to \infty$)
of a noise source in the absence of flow,
whose expected output is
$\DMLSNgenavgt{\DMLSt{}} = 1, \DMLSt{} \ge 0$, is given by
\ifOneCol
\begin{align}
\DMLSNnavg
= \frac{1}{2\DMLSk{}}\!  \left[\beta + 1 + \frac{1}{\DMLSx_n}
\EXP{-\DMLSr{sum}{\DMLSk{}}^{\frac{1}{2}}}\!\!
\left({\DMLSk{}}^{-\frac{1}{2}} + \DMLSr{obs}\right)\!
-\frac{1}{\DMLSx_n}
\EXP{-|\DMLSr{dif}|{\DMLSk{}}^{\frac{1}{2}}}\!\!
\left({\DMLSk{}}^{-\frac{1}{2}} + \beta\DMLSr{obs}\right)
\!\right]\!\!,
\label{EQ13_05_53}
\end{align}
\else
\begin{align}
\DMLSNnavg
= &\; \frac{1}{2\DMLSk{}}\bigg[1-\frac{1}{\DMLSx_n}
\EXP{-|\DMLSr{dif}|{\DMLSk{}}^{\frac{1}{2}}}\!\!
\left({\DMLSk{}}^{-\frac{1}{2}} + \beta\DMLSr{obs}\right)
\nonumber \\
&+\frac{1}{\DMLSx}
\EXP{-\DMLSr{sum}{\DMLSk{}}^{\frac{1}{2}}}\!\!
\left({\DMLSk{}}^{-\frac{1}{2}} + \DMLSr{obs}\right)+\beta
\bigg],
\label{EQ13_05_53}
\end{align}
\fi
where $\DMLSr{dif} = \DMLSr{obs}-\DMLSx_n$,
$\DMLSr{sum} = \DMLSr{obs}+\DMLSx_n$,
$\beta = \sgn(\DMLSr{obs}-\DMLSx_n)$, and $\sgn(\cdot)$ is the sign function.
\end{theorem}
\begin{IEEEproof}
Please refer to Appendix~\ref{app_no_flow}.
\end{IEEEproof}

\subsubsection{Absence of Flow and Molecule Degradation}

Time-varying solutions to both (\ref{EQ13_07_27_int}) and (\ref{EQ13_05_29})
are only possible in the absence of flow and molecule degradation, i.e., if
$\Pecpara = \Pecper = 0$ and $\DMLSk{} = 0$. If we are using the \UCA, then
(\ref{EQ13_07_27_int}) can be combined with \cite[Eq. (3.5b)]{RefWorks:586} and
we can write
\begin{equation}
\DMLSNntavg{\DMLSt{}} = \frac{\DMLSV}{4\pi\DMLSx_n}
\left(1-\ERF{\frac{\DMLSx_n}{2\sqrt{\DMLSt{}}}}\right),
\label{EQ13_05_18}
\end{equation}

\begin{remark}
\label{remark_far_no_flow}
We see from (\ref{APR12_32}) that $\ERF{a} \to 0$ as $a \to 0$,
and that  $\ERF{a} \approx 0.056$ when $a = 0.05$. If the
reference distance is chosen to be the distance of the noise
source from the receiver, i.e., $\dist = \x_n$,
and if there is no advection or molecule degradation,
then from (\ref{EQ13_05_18}) we must wait until $\DMLSt{} > 100$
before the impact is expected to be at least $95\,\%$ of the asymptotic
impact.
\end{remark}

The time-varying solution to (\ref{EQ13_05_29}) is presented in the following
theorem:

\begin{theorem}[$\DMLSNntavg{\DMLSt{}}$ in Absence of Flow and Degradation]
\label{theorem_no_flow_degrad}
The expected time-varying impact of a noise source, whose expected output is
$\DMLSNgenavgt{\DMLSt{}} = 1, \DMLSt{} \ge 0$, is given in the absence of flow
and molecule degradation by
\ifOneCol
\begin{align}
\DMLSNntavg{\DMLSt{}} = &\; \ERF{\frac{\DMLSr{dif}}{2\sqrt{\DMLSt{}}}}\!\!
\left[\frac{{\DMLSr{dif}}^2}{4} + \frac{\DMLSt{}}{2}
+ \frac{{\DMLSr{dif}}^3}{6\DMLSx_n}\right]
+ \ERF{\frac{\DMLSr{sum}}{2\sqrt{\DMLSt{}}}}\!\!
\left[\frac{{\DMLSr{sum}}^2}{4} + \frac{\DMLSt{}}{2}
- \frac{{\DMLSr{sum}}^3}{6\DMLSx_n}\right] \nonumber \\
& - \frac{\beta{\DMLSr{dif}}^2}{4} - \frac{{\DMLSr{sum}}^2}{4}
+ \frac{{\DMLSr{sum}}^3}{6\DMLSx_n} - \frac{|\DMLSr{dif}|^3}{6\DMLSx_n}
+ \sqrt{\frac{\DMLSt{}}{\pi}} \EXP{-\frac{{\DMLSr{dif}}^2}{4\DMLSt{}}}\!\!
\left[\frac{\DMLSr{dif}}{2} - \frac{2\DMLSt{}}{3\DMLSx_n}
+ \frac{{\DMLSr{dif}}^2}{3\DMLSx_n}\right] \nonumber \\
& + \sqrt{\frac{\DMLSt{}}{\pi}} \EXP{-\frac{{\DMLSr{sum}}^2}{4\DMLSt{}}}\!\!
\left[\frac{\DMLSr{sum}}{2} + \frac{2\DMLSt{}}{3\DMLSx_n}
- \frac{{\DMLSr{sum}}^2}{3\DMLSx_n}\right].
\label{EQ13_05_38}
\end{align}
\else
\begin{align}
\DMLSNntavg{\DMLSt{}} = &\; \ERF{\frac{\DMLSr{dif}}{2\sqrt{\DMLSt{}}}}\!\!
\left[\frac{{\DMLSr{dif}}^2}{4} + \frac{\DMLSt{}}{2}
+ \frac{{\DMLSr{dif}}^3}{6\DMLSx_n}\right] \nonumber \\
& + \ERF{\frac{\DMLSr{sum}}{2\sqrt{\DMLSt{}}}}\!\!
\left[\frac{{\DMLSr{sum}}^2}{4} + \frac{\DMLSt{}}{2}
- \frac{{\DMLSr{sum}}^3}{6\DMLSx_n}\right] \nonumber \\
& + \sqrt{\frac{\DMLSt{}}{\pi}} \EXP{-\frac{{\DMLSr{dif}}^2}{4\DMLSt{}}}\!\!
\left[\frac{\DMLSr{dif}}{2} - \frac{2\DMLSt{}}{3\DMLSx_n}
+ \frac{{\DMLSr{dif}}^2}{3\DMLSx_n}\right] \nonumber \\
& + \sqrt{\frac{\DMLSt{}}{\pi}} \EXP{-\frac{{\DMLSr{sum}}^2}{4\DMLSt{}}}\!\!
\left[\frac{\DMLSr{sum}}{2} + \frac{2\DMLSt{}}{3\DMLSx_n}
- \frac{{\DMLSr{sum}}^2}{3\DMLSx_n}\right] \nonumber \\
& - \frac{\beta{\DMLSr{dif}}^2}{4} - \frac{{\DMLSr{sum}}^2}{4}
+ \frac{{\DMLSr{sum}}^3}{6\DMLSx_n} - \frac{|\DMLSr{dif}|^3}{6\DMLSx_n}.
\label{EQ13_05_38}
\end{align}
\fi
\end{theorem}
\begin{IEEEproof}
Please refer to Appendix~\ref{app_no_flow_degrad}.
\end{IEEEproof}

Although (\ref{EQ13_05_38}) is verbose, it can be evaluated for \emph{any}
non-zero values of $\DMLSx_n$ and $\DMLSt{}$. We consider the case
$\DMLSx_n=0$ at the end of this subsection. Here, we note that the
asymptotic impact of a noise source, i.e., as $\DMLSt{} \to \infty$, can be
evaluated from (\ref{EQ13_05_38}) using the properties of limits and
l'H\^{o}pital's rule as
\begin{equation}
\DMLSNnavg = \frac{{\DMLSr{sum}}^3}{6\DMLSx_n} - \frac{|\DMLSr{dif}|^3}{6\DMLSx_n}
- \frac{{\DMLSr{sum}}^2}{4} - \frac{\beta{\DMLSr{dif}}^2}{4}.
\label{EQ13_05_60}
\end{equation} 

\begin{remark}
\label{remark_no_flow}
Eq. (\ref{EQ13_05_60}) simplifies to a single term if the noise source is
outside of the receiver, i.e., if $\DMLSr{dif} = \DMLSr{obs}-\DMLSx_n < 0$.
It can then be shown that $\DMLSNnavg = {\DMLSr{obs}}^3/(3\DMLSx_n)$,
which is equivalent to (\ref{EQ13_07_27}) with a spherical receiver
in the absence of advection
and molecule degradation, i.e., $\Pecpara = \Pecper = \DMLSk{} = 0$,
even though (\ref{EQ13_07_27}) was derived for a noise source that
is \emph{far} from the receiver. Thus, in the absence of advection and molecule
degradation, the expected impact of a noise source \emph{anywhere}
outside the receiver increases with the inverse of the distance to the
receiver.
\end{remark}

\subsubsection{Worst-Case Noise Source Location}

Finally, we consider the special case where the noise source is located
\emph{at} the receiver, i.e., $\DMLSx_n=0$. Clearly, the \UCA\, should not apply
in this case, so we only consider the evaluation of (\ref{EQ13_05_29}).
Generally, we need to apply l'H\^{o}pital's rule to account for $\DMLSx_n=0$,
and here we do so for three cases. First, if evaluating
(\ref{EQ13_05_29}) directly and $\Pecpara = 0$, then l'H\^{o}pital's rule must
be used to re-write the second term inside the curly braces in
(\ref{EQ13_05_29}) (i.e., the term with the two exponentials, including the
scaling by $\sqrt{\DMLStau/\pi}/\DMLSrad{eff}$) as
\begin{equation}
-\frac{\DMLSr{obs}}{(\pi\DMLStau)^\frac{1}{2}}\EXP{-\frac{{\DMLSr{obs}}^2}{4\DMLStau}}.
\label{EQ13_05_29_x0}
\end{equation}

Second, if evaluating (\ref{EQ13_05_53}), which applies asymptotically in
the absence of flow, then we can apply l'H\^{o}pital's rule
in the limit of $\DMLSx_n \to 0$ and write (\ref{EQ13_05_53}) as
\begin{equation}
\lim_{\DMLSx_n \to 0} \DMLSNnavg = \frac{1}{\DMLSk{}} -
\EXP{-\DMLSr{obs}\sqrt{\DMLSk{}}}\left(\frac{1}{\DMLSk{}} +
\frac{\DMLSr{obs}}{\sqrt{\DMLSk{}}}\right).
\label{EQ13_05_54}
\end{equation}

\begin{remark}
\label{remark_degradation}
From (\ref{EQ13_05_53}) and (\ref{EQ13_05_54}) we see that any increase
in $\DMLSk{}$ will result in a decrease in the expected number of noise
molecules observed, even if the noise source is located at the receiver
(i.e., $\DMLSx_n = 0$).
\end{remark}

Third, the time-varying impact of the ``worst-case'' noise source in the absence
of flow and molecule degradation can be found using repeated applications of
l'H\^{o}pital's rule to (\ref{EQ13_05_38}) as
\ifOneCol
\begin{equation}
\lim_{\DMLSx_n \to 0} \DMLSNntavg{\DMLSt{}} =
\ERF{\frac{\DMLSr{obs}}{2\sqrt{\DMLSt{}}}}\left[\DMLSt{} -
\frac{{\DMLSr{obs}}^2}{2}\right]
- \DMLSr{obs}\sqrt{\frac{\DMLSt{}}{\pi}}\EXP{-\frac{{\DMLSr{obs}}^2}{4\DMLSt{}}}
+ \frac{{\DMLSr{obs}}^2}{2}.
\label{EQ13_05_39}
\end{equation}
\else
\begin{align}
\lim_{\DMLSx_n \to 0} \DMLSNntavg{\DMLSt{}} = &\;
\ERF{\frac{\DMLSr{obs}}{2\sqrt{\DMLSt{}}}}\left[\DMLSt{} -
\frac{{\DMLSr{obs}}^2}{2}\right] \nonumber \\
&- \DMLSr{obs}\sqrt{\frac{\DMLSt{}}{\pi}}\EXP{-\frac{{\DMLSr{obs}}^2}{4\DMLSt{}}}
+ \frac{{\DMLSr{obs}}^2}{2}.
\label{EQ13_05_39}
\end{align}
\fi

This subsection considered a number of solutions
to (\ref{APR12_42_DMLS}), where the expected molecule emission is described
as a step function. In Table~\ref{table_noise}, we summarize precisely which
conditions and assumptions apply to each equation. We will see the accuracy
of these equations in comparison with simulated noise sources in
Section~\ref{sec_num}. In practice, these equations can enable us to
more accurately assess
the effect of noise sources on the bit error probability of the intended
communication link (as we did in \cite{RefWorks:747, RefWorks:752}, where
we only assumed that the expected impact of noise sources was known).
In the remainder of this paper, we focus
on using the noise analysis to approximate some or all of the
signal observed by transmitters that release impulses of molecules.

\ifOneCol
\else
	\tableNoise{!tb}
\fi

\section{Multiuser Interference}
\label{sec_mui}

In this section, we consider the impact of transmitters that are using
the same modulation scheme as the transmitter that is linked to
the receiver of interest but are sending independent information. Thus,
the $\A$ molecules emitted by these unintended transmitters are effectively
noise. We begin by presenting the complete model of the observations
made at the receiver due to any number of transmitters (independent of whether
the transmitters are linked to the receiver). This detailed model is
the most comprehensive, so it enables the most accurate calculation of
the bit error probability,
but it requires knowledge of all transmitter sequences. Then,
we apply our results in Section~\ref{sec_noise} to simplify the
analysis of an interfering transmitter.

\subsection{Complete Multiuser Model}

Consider from (\ref{EQ13_05_28_obs}) that all $\U$ sources of $\A$ molecules
are transmitters with the same modulation scheme.
Transmitter $u$ has independent binary sequence
$\dataSeqU{u} = \{\dataU{1}{u},\dataU{2}{u},\dots\}$ to send to its intended
receiver, where $\dataU{j}{u}$ is the $j$th information bit and
$\Pr(\dataU{j}{u} = 1) = \Pone$. The only receiver that we are concerned
with is the one at the origin. The transmitters do not coordinate
their transmissions so they all transmit simultaneously, but for
clarity of exposition we assume that the transmitters are
initially synchronized and begin transmitting at $\DMLSt{} = 0$. It
is also straightforward to add an initial timing offset to each
transmitter, but we omit that extension in this paper in
order to focus on asymptotic multiuser interference.
Transmitter $u$ has bit interval $\TU{u}$ seconds and it releases
$\NemitU{u}$ $\A$ molecules at the start of the interval to send a binary
$1$ and no molecules to send a binary $0$.
We model instantaneous molecule releases as
approximations of releases that are much shorter than the
bit interval; we do not expect that instantaneous releases are practical.
Furthermore, we define
the dimensionless bit interval $\DMLSTU{u}$ and dimensionless
number of emitted molecules $\DMLSNemitU{u}$, where we scale the
dimensional variables by $\Dx{\A}/\dist^2$ and $1/\NAx{REF}$,
respectively. We note that this binary modulation scheme can be easily
extended to any pulse amplitude modulation scheme, where the $u$th
transmitter encodes multiple bits in the number of molecules released
at one time.

The channel impulse response is the same as in the general noise source
case, i.e., (\ref{APR12_22}), where we adjust the frame of reference
for each transmitter so that it lies along the $\DMLSx$-axis.
From (\ref{APR12_42_DMLS}), we immediately have the expected number
of molecules observed due to an emission by transmitter $u$ at
time $\DMLSt{} = 0$, $\DMLSNxtavg{\DMLSt{}}{tx,u}$, written as
\begin{equation}
\label{EQ13_07_27_int_gen}
\DMLSNxtavg{\DMLSt{}}{tx,u} =
\int\limits_0^{\DMLSr{obs}}
\int\limits_{0}^{2\pi}
\int\limits_{0}^{\pi}
{\DMLSr{i}}^2\DMLSNemitU{u}
\DMLSC{\DMLSA}\sin\theta
d\theta d\phi d\DMLSr{i},
\end{equation}
which we have shown in \cite{RefWorks:662} can be accurately
approximated as a time-varying Poisson random variable when
in dimensional form.
At any moment, the distribution of molecules observed at our
intended receiver is the sum of molecules expected from all
emissions made by all transmitters, as given
in (\ref{EQ13_05_28_obs}). This is (dimensionally)
a Poisson random
variable (because it is a sum of independent Poisson random
variables; see \cite[Ch. 5.2]{RefWorks:725}) that has
(dimensionless) mean
\begin{equation}
\label{EQ13_08_02}
\DMLSNxtavg{\DMLSt{}}{obs} =
\sum_{u=1}^{\U}
\sum_{j=1}^{\floor{\frac{\DMLSt{}}{\DMLSTU{u}}+1}}\!\!\!\!
\dataU{j}{u}\DMLSNxtavg{\DMLSt{}-(j-1)\DMLSTU{u}}{tx,u}\!.
\end{equation}

\subsection{Asymptotic Interference}

Precise analysis of the performance of the receiver's detector
can be made using (\ref{EQ13_08_02}), but we must have knowledge
of every transmitter sequence $\dataSeqU{u}$. We propose simplifying
the analysis by applying our results in Section~\ref{sec_noise}.
For widest applicability, i.e., to include molecule degradation
and flow in any direction, we assume that interfering transmitters
are sufficiently far away to apply the uniform concentration
assumption (this makes sense; an interferer that is very close to
the receiver would likely result in an error probability that
is too high for communication with the intended
transmitter to be practical). The corresponding closed-form analysis is
asymptotic in time, but this is acceptable because we can assume
that interferers were transmitting for a long time before the start
of our intended transmission (we will see in Section~\ref{sec_num}
that this is an easy assumption to satisfy).
The remainder of this section
can also be easily extended to the other special cases in
Section~\ref{sec_noise}. 

Consider the asymptotic impact of a single interfering transmitter.
The emissions of the $u$th
transmitter must be approximated as a continuous function so that we
can apply the results from our noise analysis. The effective
emission rate is $\Pone\DMLSNemitU{u}$ molecules every $\DMLSTU{u}$
dimensionless time units. If we choose
$\NAx{REF} = \dist^2\Pone\NemitU{u}/(\TU{u}\Dx{\A})$,
then the emission function of the $u$th transmitter can be approximated as
$\DMLSNgenavgt{\DMLSt{}} = 1, \DMLSt{} \ge 0$. From (\ref{EQ13_07_27}),
we immediately have the expected asymptotic impact of the noise source,
$\DMLSNxavg{u}$, written as
\begin{equation}
\DMLSNxavg{u}
= \frac{\DMLSV}{4\pi\DMLSx_u}
\EXP{\frac{\DMLSx_u\Pecpara}{2}
- \frac{\DMLSx_u}{2}\sqrt{{\Pecpara}^2 + {\Pecper}^2
+4\DMLSk{}}},
\label{EQ13_08_05}
\end{equation}
where we recall that, in general, we have adjusted the reference coordinate
frame so that the $u$th transmitter lies on the negative $\DMLSx$-axis.
The time-varying impact
can be found via numerical integration of (\ref{EQ13_07_27_int}), or, if
$\Pecpara = \Pecper = 0$ and $\DMLSk{} = 0$, i.e., if there is no
advection or molecule degradation, via (\ref{EQ13_05_18}).
The complete asymptotic multiuser interference is given by adding
(\ref{EQ13_08_05}) for all $\U-1$ interfering transmitters.

We note that (\ref{EQ13_08_05}) is a constant approximation of what is in practice
a signal that is expected to oscillate over time. The channel impulse
response given
by (\ref{APR12_22}) has a definitive peak and tail. The interference
can be envisioned as the most recent peak followed by all
of the tails of prior transmissions. Even asymptotically, the expected
impact at a given instant will depend on the time relative to the interferer's
transmission intervals. So, over time, (\ref{EQ13_08_05}) will both
overestimate and underestimate the impact of the interferer. However,
we expect that, on average, (\ref{EQ13_08_05}) will tend to overestimate the impact
more often. This is because the approximation of molecule emission as a
continuous function effectively makes the release of molecules later than they
actually are by ``spreading'' emissions over the entire bit interval instead
of releasing all of them at the start of the bit interval.
We will visualize the
accuracy of (\ref{EQ13_08_05}) more clearly in Section~\ref{sec_num}.

\section{Asymptotic ISI}
\label{sec_isi}

In this section, we focus on characterizing the signal observed at the
receiver due to the intended transmitter \emph{only}. We seek a method to
model some of the \ISI\, asymptotically based on the previous
analysis in this paper. Specifically, we model $\VAmem$
prior bits \emph{explicitly} (and not as a signal from a
continuously-emitting source), and the impact of all earlier bits is
approximated asymptotically as a continuously-emitting source.
The choice of $\VAmem$ enables a tradeoff between accuracy and
computational efficiency. We describe the application of our
model for asymptotic old \ISI\, to simplify the
evaluation of the expected bit error probability of weighted sum detectors,
which in general requires finding the expected probability of error of all
possible transmitter sequences and taking an average. Other applications of
our model for asymptotic \ISI\, are simplifying the implementation of
the maximum likelihood detector and in the design of a weighted sum detector
with adaptive weights.
Adaptive weighting is physically realizable in biological systems;
neurons sum inputs from synapses with dynamic weights
(see \cite[Ch. 12]{RefWorks:587}), but we leave the design of
adaptive weighted sum detectors for future work.

\subsection{Decomposition of Received Signal}

We now decompose the signal from the intended transmitter, i.e.,
$\DMLSNxt{\DMLSt{}}{1}$ in (\ref{EQ13_05_28_obs}).
To emphasize that source $1$ is the intended transmitter,
we re-write its signal as $\DMLSNxt{\DMLSt{}}{1} = \DMLSNxt{\DMLSt{}}{tx}$
and also drop the $1$ subscript from its transmission parameters. If we
apply the uniform concentration assumption for clarity of exposition, then
the expected number of observed molecules due to the intended transmitter,
$\DMLSNxtavg{\DMLSt{}}{tx}$, given the transmitter sequence $\dataSeq$, is
\begin{equation}
\label{EQ13_04_02_DMLS}
\DMLSNxtavg{\DMLSt{}}{tx} = \DMLSNemit\DMLSV
\sum_{j=1}^{\floor{\frac{\DMLSt{}}{\DMLST}+1}}
\data{j}\DMLSCxFun{\DMLSA}{\DMLSr{eff}(j)}{\DMLStau(j)},
\end{equation}
where here $({\DMLSr{eff}}(j))^2 = (\DMLSx_1 - \Pecpara\DMLStau(j))^2 +
(\Pecper\DMLStau(j))^2$ and
$\DMLStau(j) = \DMLSt{}-(j-1)\DMLST$, i.e., $({\DMLSr{eff}}(j))^2$
is the square of the effective distance between the receiver and
the transmitter's $j$th emission and $\DMLStau(j)$ is the
time elapsed since the beginning of the $j$th bit interval. For
compactness in the remainder of this section, we write
$\DMLSCxFun{\DMLSA}{\DMLSr{eff}(j)}{\DMLStau(j)} = \DMLSC{\DMLSA}(j)$. We also
note that it is straightforward to relax the uniform concentration assumption
and re-write (\ref{EQ13_04_02_DMLS}) into a more general form, following
our analysis in \cite{RefWorks:752}.
Similarly to the discussion of $\DMLSNxt{\DMLSt{}}{obs}$ in
Section~\ref{sec_mui}, $\DMLSNxt{\DMLSt{}}{tx}$ is (dimensionally)
a Poisson
random variable with time-varying mean $\DMLSNxtavg{\DMLSt{}}{tx}$.

We decompose (\ref{EQ13_04_02_DMLS}) into three terms: molecules
observed due to the current bit interval, $\DMLSNxt{\DMLSt{}}{tx,cur}$,
molecules observed that were released within $\VAmem$ intervals before
the current interval, $\DMLSNxt{\DMLSt{}}{tx,isi}$, and molecules
observed that were released in any older bit interval,
$\DMLSNxt{\DMLSt{}}{old}$. Specifically, (\ref{EQ13_04_02_DMLS})
becomes
\ifOneCol
\begin{align}
\label{EQ13_08_08_DMLS}
\DMLSNxtavg{\DMLSt{}}{tx} = &\; \DMLSNemit\DMLSV\left[
\data{j_c}\DMLSC{\DMLSA}(j_c)
+ \sum_{j=j_c-\VAmem}^{j_c-1}\data{j}\DMLSC{\DMLSA}(j)
+ \sum_{j=1}^{j_c-\VAmem-1}\data{j}\DMLSC{\DMLSA}(j)\right] \\
= &\; \DMLSNxtavg{\DMLSt{}}{tx,cur} + \DMLSNxtavg{\DMLSt{}}{tx,isi}
+ \DMLSNxtavg{\DMLSt{}}{old},
\label{EQ13_08_08_summary}
\end{align}
\else
\begin{align}
\DMLSNxtavg{\DMLSt{}}{tx} = &\; \DMLSNemit\DMLSV\Bigg[
\data{j_c}\DMLSC{\DMLSA}(j_c)
+ \sum_{j=j_c-\VAmem}^{j_c-1}\data{j}\DMLSC{\DMLSA}(j) \nonumber\\
\label{EQ13_08_08_DMLS}
& + \sum_{j=1}^{j_c-\VAmem-1}\data{j}\DMLSC{\DMLSA}(j)\Bigg] \\
= &\; \DMLSNxtavg{\DMLSt{}}{tx,cur} + \DMLSNxtavg{\DMLSt{}}{tx,isi}
+ \DMLSNxtavg{\DMLSt{}}{old},
\label{EQ13_08_08_summary}
\end{align}
\fi
where $j_c = \floor{\frac{\DMLSt{}}{\DMLST}+1}$
is the index of the \emph{current} bit interval, and we
emphasize that each term in (\ref{EQ13_08_08_summary}) is evaluated given the
current transmitter sequence $\dataSeq$.
The decomposition enables us to simplify the expression for the
signal observed due to molecules released by the transmitter if
we can write an \emph{asymptotic} expression for the expected value of
$\DMLSNxtavg{\DMLSt{}}{old}$,
i.e., $\DMLSNxavg{old}$. However, the analysis that we have derived
in this paper for the asymptotic impact of signals is dependent on
the on-going emission of molecules that began at time
$\DMLSt{}=0$. We present two methods to derive
$\DMLSNxavg{old}$. First, we begin with the asymptotic expression in
(\ref{EQ13_08_05}) for an interferer that is always emitting and then subtract
the \emph{unconditional} expected impact of molecules released within the last
$\VAmem+1$ bit intervals. We then write the expected and
\emph{time-varying} but \emph{asymptotic}
expression for $\DMLSNxtavg{\DMLSt{}}{old}$ as
\begin{equation}
\DMLSNxtavg{\DMLSt{}}{old} = \DMLSNxavg{tx}
- \DMLSNxtavg{\DMLSt{}}{tx,isi} - \DMLSNxtavg{\DMLSt{}}{tx,cur},
\label{EQ13_08_09}
\end{equation}
where $\DMLSNxavg{tx}$ is in the same form as (\ref{EQ13_08_05}),
and here $\DMLSNxtavg{\DMLSt{}}{tx,cur}$ and
$\DMLSNxtavg{\DMLSt{}}{tx,isi}$ do \emph{not} depend on $\dataSeq$ because
they are averaged over the 2 and $2^{\VAmem}$ possible corresponding
bit sequences, respectively. Eq.~(\ref{EQ13_08_09}) is time-varying
because the expected impact that we subtract depends on the time within the
current bit interval.
From (\ref{EQ13_08_09}), $\DMLSNxt{\DMLSt{}}{old}$ is asymptotically a
cyclostationary process; the expected mean is periodic
with period $\DMLST$. Although (\ref{EQ13_08_09}) is tractable, it
is cumbersome to evaluate (because we need to average the expected
impact of molecules released over all $2^{\VAmem+1}$ possible recent
bit sequences, including the current bit)
and is also not accurate (because (\ref{EQ13_08_05})
is based on continuous emission while
$\DMLSNxtavg{\DMLSt{}}{tx,isi}$ and $\DMLSNxtavg{\DMLSt{}}{tx,cur}$
are based on discrete emissions at the start of the corresponding bit
intervals). A second method to derive $\DMLSNxavg{old}$, which
requires less approximation, is to start with (\ref{EQ13_07_27_int}) and
change the limits of integration over time, i.e.,
\begin{equation}
\DMLSNxtavg{\DMLSt{}}{old} =
\int\limits_{\DMLSt{} - (j_c-\VAmem-1)\DMLST}^{\infty}
\DMLSV \DMLSCxFun{\DMLSA}{\DMLSr{eff}}{\DMLStau}
d\DMLStau,
\label{EQ13_08_09_int}
\end{equation}
where ${\DMLSr{eff}}^2 = (\DMLSx_1 - \Pecpara\DMLStau)^2 +
(\Pecper\DMLStau)^2$ and, if $j_c-\VAmem-1 \le 0$,
then we do not yet have asymptotic \ISI.
Eq. (\ref{EQ13_08_09_int}) is also periodic with period $\DMLST$. 
A special case of (\ref{EQ13_08_09_int}) occurs if we have $\Pecpara = \Pecper = 0$
and $\DMLSk{} = 0$. In such an environment, we can subtract
(\ref{EQ13_05_18}) from the asymptotic expression in (\ref{EQ13_07_27}).
Specifically, we can write
\begin{equation}
\DMLSNxtavg{\DMLSt{}}{old} =
\frac{\DMLSV}{4\pi\DMLSx_1}
\ERF{\frac{\DMLSx_1}{2\sqrt{\DMLSt{} - (j_c-\VAmem-1)\DMLST}}}.
\label{EQ13_05_18_isi}
\end{equation}

Depending on the environmental parameters and whether there is a preference
for tractability or accuracy, either (\ref{EQ13_08_09}), (\ref{EQ13_08_09_int}),
or (\ref{EQ13_05_18_isi}) can be used for $\DMLSNxtavg{\DMLSt{}}{old}$
in (\ref{EQ13_08_08_summary}). This asymptotic \ISI\, term is
independent of the actual transmitter data sequence $\dataSeq$, so it can be
pre-computed and used to assist in applications such as evaluating the expected
bit error probability of a weighted sum detector.

\subsection{Weighted Sum Detection}

We focus on a single type of detector at the receiver as a detailed example
of the application of an asymptotic model of old \ISI. We
proposed the family of weighted sum detectors in \cite{RefWorks:747} as
detectors that can operate with limited memory and computational requirements.
We envision such detectors to be physically practical
because they can already be found in biological systems such as neurons;
see \cite[Ch. 12]{RefWorks:587}. Here, we consider weighted sum detectors
where the receiver makes $\M$ observations in every
bit interval, and we assume that these observations are equally
spaced such that the $\smM$th observation in the $j$th interval
is made at time $\DMLSt{}(j,\smM) = \left(j+\frac{\smM}{\M}\right)\DMLST$,
where $j = \{1,2,\ldots,\B{}\}, \smM = \{1,2,\ldots,\M\}$.

The dimensionless decision rule of the weighted sum detector in
the $j$th bit interval is
\begin{equation}
\dataObs{j} = \left\{
 \begin{array}{rl}
  1 & \text{if} \quad
  \sum_{\smM = 1}^{\M}\weight{\smM}\DMLSNxt{\DMLSt{}(j,\smM)}{obs}
  \ge \DMLSthresh,\\
  0 & \text{otherwise},
 \end{array} \right.
\label{FEB13_33}
\end{equation}
where $\DMLSNxt{\DMLSt{}(j,\smM)}{obs}$ is the $\smM$th observation 
as given by (\ref{EQ13_05_28_obs}),
$\weight{\smM}$ is the weight of the $\smM$th observation, and
$\DMLSthresh$ is the binary decision threshold (we note that we
do not need to make the weights dimensionless because they already are).
We assume that a constant optimal $\DMLSthresh$ for the given environment
(and for the given formulation of \ISI\, when evaluating the expected
performance)
is found via numerical search.

Given a particular transmitter sequence $\dataSeq$, we can calculate the
expected error probability of a weighted sum detector. The
expected probability of error of the $j$th bit,
$\Pe{j | \dataSeq}$, is
\ifOneCol
\begin{equation}
 \Pe{j | \dataSeq} =
 \quad\left\{
 \begin{array}{rl}
  \Pr\left(\sum_{\smM = 1}^{\M}\weight{\smM}\DMLSNxt{\DMLSt{}(j,\smM)}{obs}
  < \DMLSthresh \right)
  & \text{if} \;\data{j} = 1,
  \vspace*{2mm}\\
  \Pr\left(\sum_{\smM = 1}^{\M}\weight{\smM}\DMLSNxt{\DMLSt{}(j,\smM)}{obs}
  \ge \DMLSthresh \right)
  & \text{if} \;\data{j} = 0.
 \end{array} \right.
 \label{EQ13_06_15}
\end{equation}
\else
\begin{align}
& \Pe{j | \dataSeq} = \nonumber \\
& \quad\left\{
 \begin{array}{rl}
  \!\Pr\left(\sum_{\smM = 1}^{\M}\weight{\smM}\DMLSNxt{\DMLSt{}(j,\smM)}{obs}
  < \DMLSthresh \right)
  & \text{if} \;\data{j} = 1,
  \vspace*{2mm}\\
  \!\Pr\left(\sum_{\smM = 1}^{\M}\weight{\smM}\DMLSNxt{\DMLSt{}(j,\smM)}{obs}
  \ge \DMLSthresh \right)
  & \text{if} \;\data{j} = 0.
 \end{array} \right.
 \label{EQ13_06_15}
\end{align}
\fi

In our previous work in \cite{RefWorks:747}, we approximated the expected error
probability for the $j$th bit averaged over all possible
transmitter sequences, $\Peavg{j}$, by averaging
(\ref{EQ13_06_15}) over a subset of all
sequences. An error probability was determined for all $\B{}$
bit intervals of every considered sequence. This analysis can be greatly
simplified by evaluating the probability of error of a single bit that
is sufficiently ``far'' from the start of the sequence, i.e., $j \to \infty$,
and then
model only the most recent $\VAmem$ intervals of \ISI\, explicitly
and represent all older intervals with $\DMLSNxtavg{\DMLSt{}}{old}$.
Furthermore, if the impacts of the external noise sources in (\ref{EQ13_05_28_obs})
are represented asymptotically (whether they are interferers or other noise sources),
or if there are no external noise sources present,
then we only need to evaluate the expected probability of error
of the \emph{last} bit in $2^{\VAmem+1}$ sequences.

The evaluation of (\ref{EQ13_06_15}) depends on the statistics of the
weighted sum $\sum_{\smM = 1}^{\M}\weight{\smM}\DMLSNxt{\DMLSt{}(j,\smM)}{obs}$.
For simplicity, we limit our discussion to the special case where the
weights are all equal, i.e., $\weight{\smM} = 1\, \forall \smM$, such
that we can assume that the (dimensional) observations are independent Poisson
random variables (we also considered the general case, where we must
approximate the observations as Gaussian random variables, in \cite{RefWorks:747}).
Then, the sum of observations is also a Poisson random variable.
The \CDF\, of the weighted sum in the
$j$th bit interval is then \cite[Eq. 38]{RefWorks:747}
\ifOneCol
\begin{equation}
\Pr\left(\sum_{\smM = 1}^{\M}\Nobst{\tx{j,\smM}} < \thresh \right) =
\EXP{-\sum_{\smM = 1}^{\M}\Nobsavg{\tx{j,\smM}}}
\sum_{i=0}^{\thresh-1}
\frac{\left(\sum\limits_{\smM = 1}^{\M}\Nobsavg{\tx{j,\smM}}\right)^i}{i!},
\label{EQ13_06_02}
\end{equation}
\else
\begin{align}
& \Pr\left(\sum_{\smM = 1}^{\M}\Nobst{\tx{j,\smM}} < \thresh\right) = \nonumber \\
& \qquad\EXP{-\sum_{\smM = 1}^{\M}\Nobsavg{\tx{j,\smM}}} \nonumber \\
& \qquad\times \sum_{i=0}^{\thresh-1}
\frac{\left(\sum\limits_{\smM = 1}^{\M}\Nobsavg{\tx{j,\smM}}\right)^i}{i!},
\label{EQ13_06_02}
\end{align}
\fi
where, from (\ref{EQ13_05_28_obs}),
\begin{equation}
\Nobsavg{\tx{j,\smM}} = \Ntxtavg{\tx{j,\smM}} +
\sum_{u=2}^{\U}\NAxavg{u},
\label{EQ13_05_28_obs_avg}
\end{equation}
and $\Ntxtavg{t}$ and $\NAxavg{u}$ are the \emph{dimensional} forms
of the number of molecules expected from the intended transmitter and $u$th
noise source, i.e.,
$\DMLSNxtavg{\DMLSt{}}{tx}$ of $\DMLSNxavg{u}$, respectively, and we
emphasize that we represent the noise sources asymptotically.
We write (\ref{EQ13_06_02}) and (\ref{EQ13_05_28_obs_avg})
in dimensional form
to emphasize that the observations are discrete.
For the corresponding simulations in Section~\ref{sec_num}, we only
consider $\U=1$ to focus on the accuracy of the asymptotic approximation
of old \ISI, and we evaluate the
old \ISI\, as given by (\ref{EQ13_08_09_int}) or (\ref{EQ13_05_18_isi}) for
$\kth{}\ne 0$ and $\kth{}=0$, respectively.

\section{Numerical Results}
\label{sec_num}

In this section, we present numerical and simulation results to verify
the analysis of noise, multiuser interference, and \ISI\, performed
in this paper.
To clearly show the accuracy of all equations derived in this paper,
we simulate only \emph{one} source at a time,
measuring either 1) the impact of a noise source or an interfering transmitter,
or 2) the receiver error probability when the intended transmitter is the only
molecule source.
Our simulations are executed in the particle-based stochastic
framework that we introduced in \cite{RefWorks:631, RefWorks:662}.
The $\A$ molecules are initialized at the corresponding source
when they are released. The location of each molecule, as determined by
the uniform flow and random diffusion, is updated every time step $\Delta t$,
where diffusion along each dimension is simulated by generating a normal random
variable with variance $2\Dx{\A} \Delta t$. If
there is molecule degradation, then every molecule has a chance of
degrading in every time step with probability $\kth{}\Delta t$.
If there is no molecule degradation, then all molecules released are present
indefinitely. The
signal at the receiver is updated in every time step by counting the
number of $\A$ molecules that are within $\robs$ of the origin.

Constant environmental parameters are listed in Table~\ref{table_param}.
The chosen values are consistent with those that we considered in
\cite{RefWorks:752}, where we noted that the value of the diffusion
coefficient $\Dx{\A}$ is similar to that
of many small molecules in water at room temperature
(see \cite[Ch. 5]{RefWorks:742}), and is also comparable to
that of small biomolecules in blood plasma (see \cite{RefWorks:754}).
Most of the results in this section have been non-dimensionalized with
the reference distance $\dist$ depending on the distance from the source
of molecules to the receiver. For reference, conversions between the
dimensional variables that were simulated and their values in dimensionless
form are listed in Table~\ref{table_dmls}.

\ifOneCol
\else
	\tableParam{!tb}
\fi

\ifOneCol
\else
	\tableDMLS{!tb}
\fi

\subsection{Continuous Noise Source}

We first present the time-varying impact of the continuously-emitting
noise source that we analyzed in Section~\ref{sec_noise}.
The times between the release of consecutive molecules from the noise source
are simulated as a continuous Poisson process so that the times between
molecule release are independent. The expected release rate,
$1.2\times10^6\,\frac{\molecule}{\second}$, is chosen so that,
asymptotically, one (dimensional) molecule is expected to be observed
at the receiver due to a noise source placed $50\,$nm from the center
of the receiver (this distance is actually at the edge of the receiver,
cf. Table~\ref{table_param}).
To accommodate the range of distances considered, we adjust the
simulation time step $\Delta t$
so that $10$ steps are made within every $\DMLSt{} = 1$
time unit. Simulations are averaged over $10^5$ independent realizations.
The specific equations used for calculating the expected values, both
time-varying and asymptotically, were chosen as appropriate from
Table~\ref{table_noise}.

In Fig.~\ref{fig_noise_v0_k0}, we show the time-varying impact of the noise source
when there is no advection and no molecule degradation, i.e.,
$\Pecpara = \Pecper = 0$ and $\DMLSk{} = 0$.
Under these conditions,
we have the expected time-varying and asymptotic impact in closed form.
For every distance shown,
the impact approaches the asymptotic value as $\DMLSt{} \to 100$,
as expected from Remark \ref{remark_far_no_flow}.
The expected impact without the \UCA\, is highly accurate for all time,
and the expected impact with the \UCA\, shows visible deviation only
for $\DMLSt{} < 1$
when $\x_n < 200\,$nm, i.e., when the noise source is not far from the
receiver. We also observe that the overall impact decreases as the noise
source is placed further from the receiver; doubling the distance decreases
$\DMLSNntavg{\DMLSt{}}$ by about a factor of $8$ while
the corresponding value of $\NAx{REF}$,
defined as $\NAx{REF} = \dist^2\Ngen/\Dx{\A}$ and
used to convert
$\DMLSNntavg{\DMLSt{}}$ into dimensional form, only increases by a factor of
$4$ (see Table~\ref{table_dmls}). The overall (dimensional) decrease
in impact by a factor of $2$ is as expected by Remark \ref{remark_no_flow}.

\ifOneCol
\else
	\figNoiseNoVK{!tb}{1}
\fi

In Fig.~\ref{fig_noise_v0_k1}, we consider the same environment as
in Fig.~\ref{fig_noise_v0_k0}
but we set the molecule degradation rate $\DMLSk{} = 1$.
The accuracy of the expected expressions is comparable to
that observed in Fig.~\ref{fig_noise_v0_k0}, but here the asymptotic impact is
approached about two orders of magnitude faster, as $\DMLSt{} \to 2$.
The asymptotic impact at any distance is also less than half of that
observed in Fig.~\ref{fig_noise_v0_k0} because of the molecule
degradation.

\ifOneCol
\else
	\figNoiseKNoV{!tb}{1}
\fi

In Fig.~\ref{fig_noise_k}, we observe the impact of a noise source at the
``worst-case'' location, i.e., $\x_n = 0$, and we vary the molecule
degradation rate $\DMLSk{}$.
The expressions for the expected time-varying and asymptotic impact are
both highly accurate. We see the general trend that the
asymptotic impact decreases (as expected by Remark \ref{remark_degradation})
and is reached sooner as $\DMLSk{}$ increases.
Increasing $\DMLSk{}$ also degrades the signal from the desired
transmitter, but this can be good for reducing \ISI\, as we will see in the following
subsection.
Furthermore, it is interesting that the impact of the noise source
can be significantly reduced by increasing the rate of noise molecule
degradation, even though the noise molecules are being emitted directly
at the receiver. This implies that, if they were not degraded,
significantly more noise molecules
would have been observed by the receiver before diffusing away.

\ifOneCol
\else
	\figNoiseK{!tb}{1}
\fi

In Figs.~\ref{fig_noise_v1} and \ref{fig_noise_v1_far}, we consider the
effect of advection
on the impact of noise without molecule degradation. For clarity, we
observe $\x_n = \{0,100\}\,$nm in Fig.~\ref{fig_noise_v1} and
$\x_n = \{200,400\}\,$nm in Fig.~\ref{fig_noise_v1_far}.
When $\x_n = 0$, only one flow direction is relevant
because all flows are equivalent by symmetry. As with molecule degradation,
we observe that the presence of advection reduces the time required for
the impact of the noise source to become asymptotic, which here occurs by about
$\DMLSt{} = 4$. Flows that are not in the direction of a line from the noise
source to the receiver, i.e., $\Pecpara < 0$ or $\Pecper \neq 0$
(which we termed ``disruptive'' flows in
\cite{RefWorks:752}), decrease the asymptotic impact of the noise source.
However, the flow $\Pecpara = 1$ results in about the same
asymptotic impact as the no-flow case when $\x_n \ne 0\,$nm,
which we expect from Remark \ref{remark_far_flow},
although it might not be an intuitive result.

\ifOneCol
\else
	\figNoiseV{!tb}{1}
\fi

\ifOneCol
\else
	\figNoiseVFar{!tb}{1}
\fi

\subsection{Interference and \ISI}

We now assess the accuracy of approximating
transmitters as continuously-emitting noise sources.
First, we observe the impact of an interfering transmitter.
Second, we assess the accuracy of evaluating the receiver error probability where
we vary the number $\VAmem$ of symbols of \ISI\, treated explicitly and
approximate all older \ISI\, as an asymptotic noise source.
We consider transmitters with a common set of
dimensional transmission parameters, as described in
Table~\ref{table_param}.

In Fig.~\ref{fig_interferer}, we show the time-varying impact on the
receiver of a single interferer
using binary-encoded impulse modulation,
both with and without molecule degradation, for the interferer
placed $\x_2 = 400\,$nm or $1\,\mu$m from the receiver
(we emphasize that the \emph{only} active molecule source
is \emph{not} the intended transmitter by using the subscript $2$).
At both distances, the same bit interval is used ($\TU{2} = 0.2\,$ms).
The expected time-varying
and asymptotic curves are evaluating using (\ref{EQ13_07_27_int}) and
(\ref{EQ13_08_05}), respectively. The simulations are averaged over $10^5$
independent realizations, and in Fig.~\ref{fig_interferer} we clearly observe
oscillations in the simulated values above and below the expected curves.
The relative amplitude of these oscillations is much greater when the interferer
is closer to the receiver, and also greater when there is molecule
degradation; when $\x_2 = 400\,$nm and $\DMLSk{} = 1$, the impact in the
asymptotic regime varies from $4\times10^{-5}$ to over $6\times10^{-4}$,
but when $\x_2 = 1\,\mu$m and $\DMLSk{} = 0$, the relative amplitude of
the oscillations
is an order of magnitude smaller.
Thus, the impact of an interferer
that is sufficiently far from the receiver can be accurately approximated
with a non-oscillating function,
and an interferer does not need to be transmitting for a very long time
to assume that its impact is asymptotic ($8$ and $50$ bit intervals
are shown in Fig.~\ref{fig_interferer} for the interferers at $400\,$nm and
$1\,\mu$m, respectively; the difference is due to plotting on a
dimensionless time axis).
We note that the relative amplitude of oscillations would also decrease
if the interferer transmitted with a smaller bit interval.

\ifOneCol
\else
	\figInterferer{!tb}{1}
\fi

In Fig.~\ref{fig_isi}, we measure the average bit error probability of
the equal weight detector when $\M = 10$ samples are taken per bit interval
and the optimal decision threshold is found numerically. The receiver is placed
$\x_1 = 400\,$nm from the transmitter and we vary $\DMLSk{}$ to control
the amount of \ISI\, that we expect (since a faster molecule degradation
rate means that emitted molecules are less likely to exist sufficiently long to
interfere with future transmissions). We do not add
any external noise or interference (i.e., there is only \emph{one} source of
information molecules), but we vary the number $\VAmem$ of
bit intervals that are treated explicitly as \ISI, i.e., the complexity of
$\DMLSNxtavg{\DMLSt{}}{tx,isi}$, in evaluating the expected error probability.
Simulations are averaged over $10^4$ independent realizations, and we ignore
the decisions made within the first $50$ of the $100$ bits in each sequence
in order to approximate the ``old'' \ISI\, as asymptotic. The old
\ISI, $\DMLSNxt{\DMLSt{}}{old}$,
is found by evaluating (\ref{EQ13_08_09_int}) (or by using
(\ref{EQ13_05_18_isi}) when $\DMLSk{} = 0$), and to emphasize the
benefit of including this term we also consider evaluating the expected error
probability where we set $\DMLSNxtavg{\DMLSt{}}{old} = 0$.

\ifOneCol
\else
	\figISI{!tb}{1}
\fi

We generally observe in Fig.~\ref{fig_isi} that, as $\VAmem$ increases, the
expected error probability becomes more accurate because we treat more
of the \ISI\, explicitly instead of as asymptotic noise via
$\DMLSNxtavg{\DMLSt{}}{old}$. The
exception to this is when $\DMLSk{}=0$ and we calculate the expected value
using $\DMLSNxtavg{\DMLSt{}}{old}$. The \ISI\, in that case is much greater
than when $\DMLSk{} > 0$, such that the expected bit error probability is more
sensitive to the approximation for $\DMLSNxtavg{\DMLSt{}}{old}$, which assumes
that the release of molecules is continuous over the entire bit interval. This
approximation means that the expected ``old'' \ISI\, is overestimated and a
higher expected bit error probability is calculated. When $\DMLSk{} > 0$, the
expected bit error probability tends to underestimate that observed via
simulation because the evaluation of the expected bit error probability
assumes that all observed samples are independent, but this assumption
loses accuracy for larger $\M$. Importantly, the expected bit error
probability tends to that observed via simulation much faster when including
$\DMLSNxtavg{\DMLSt{}}{old}$, even though it is an approximation. For all
values of $\DMLSk{}$ considered, it is sufficient to consider only $2$ or $3$
intervals of \ISI\, explicitly while approximating all prior intervals
with $\DMLSNxtavg{\DMLSt{}}{old}$. If we use $\DMLSNxtavg{\DMLSt{}}{old} = 0$,
as is common in the existing literature,
then many more intervals of explicit \ISI\, are needed for comparable accuracy
($\VAmem = 20$ is still
not sufficient if $\DMLSk{} = 0$, although $\VAmem = 5$ might be acceptable
if $\DMLSk{} = 0.2$). Since the computational complexity of evaluating the
expected bit error probability increases exponentially with $\VAmem$ (because
we need to evaluate the expected probability of error due to all $2^{\VAmem+1}$
bit sequences), approximating old \ISI\, with
$\DMLSNxtavg{\DMLSt{}}{old}$ provides an effective means with which to reduce
the complexity without making a significant sacrifice in accuracy.

\section{Conclusion}
\label{sec_concl}

In this paper, we proposed a unifying model to account for the observation
of unintended molecules by a passive receiver in a diffusive molecular
communication system, where the unintended molecules include those emitted
by the intended transmitter in previous bit intervals, those emitted by
interfering transmitters, and those emitted by other external noise sources
that are continuously emitting molecules. We presented the general
time-varying expression for the expected impact of a noise source that is
emitting continuously, and then we considered a series of
special cases that facilitate time-varying or asymptotic solutions.
Knowing the expected impact of noise sources enables us to find the
effect of those sources on the bit error probability of a communication link.
We used the analysis for asymptotic
noise to approximate the impact of an interfering transmitter,
which we extended to
the general case of multiuser interference. Finally, we
decomposed the signal received from the intended transmitter so
that we could approximate
``old'' \ISI\, as asymptotic interference.
We showed how this
approximation could be used to simplify the evaluation of the expected bit
error probability of a weighted sum detector. Our simulation results
showed the high accuracy of our expressions for time-varying and
asymptotic noise. We showed that an interfering transmitter placed sufficiently
far from the receiver can be approximated as an asymptotic noise source
soon after it begins transmitting, and that approximating old \ISI\, as
asymptotic noise is an effective method to reduce the computational complexity
of evaluating the expected probability of error. Our future work includes
investigating the expected impact of noise sources with random locations,
which can be used to model the random generation of noise molecules anywhere
in the propagation medium,
and using the approximation for asymptotic \ISI\, to design adaptive
detectors, where the decision threshold is adjusted based on the knowledge of
the previously received information.

\appendix

\subsection{Proof of Theorem 1}
\label{app_no_flow}

The asymptotic integration (i.e, as $\DMLSt{} \to \infty$)
in (\ref{EQ13_05_29}) to prove
Theorem~\ref{theorem_no_flow} can be written as the summation
of four integrals which can be found by solving the following
two integrals:
\begin{align}
\label{EQ13_05_45}
& \int\limits_0^\infty \ERF{\frac{a}{{\DMLStau}^\frac{1}{2}}}
\EXP{-\DMLSk{}\DMLStau}d\DMLStau, \\
\label{EQ13_05_46}
& \int\limits_0^\infty {\DMLStau}^\frac{1}{2}
\EXP{-\frac{b}{\DMLStau}-\DMLSk{}\DMLStau}d\DMLStau,
\end{align}
where $a$ could be positive or negative and the latter occurs
only when $\DMLSx_n > \DMLSr{obs}$. To solve (\ref{EQ13_05_45}) for
$a > 0$,
we apply the substitution $c = a/{\DMLStau}^\frac{1}{2}$ and
use the definite integral \cite[Eq. 4.3.28]{RefWorks:700}
\begin{equation}
\int\limits_0^\infty \ERF{b_1c}
\EXP{-\frac{b_2^2}{4c^2}}\frac{dc}{c^3} = \frac{2}{b_2^2}
\left(1-\EXP{-b_1b_2}\right),
\end{equation}
so that we can write (\ref{EQ13_05_45}) as
\begin{equation}
\label{EQ13_05_47}
\frac{1}{\DMLSk{}}\left(1-\EXP{-2a\sqrt{\DMLSk{}}}\right).
\end{equation}

Recalling that $\ERF{\cdot}$ is an odd function,
i.e., $\ERF{-c} = -\ERF{c}$, we solve (\ref{EQ13_05_45})
for $a < 0$ as
\begin{equation}
\label{EQ13_05_49}
\frac{1}{\DMLSk{}}\left(\EXP{2a\sqrt{\DMLSk{}}}-1\right).
\end{equation}

We can solve (\ref{EQ13_05_46}) by directly applying
\cite[Eq. 3.471.16]{RefWorks:402} as
\begin{equation}
\label{EQ13_05_52}
\frac{\sqrt{\pi}}{2\DMLSk{}}\EXP{-2\sqrt{b\DMLSk{}}}
\left({\DMLSk{}}^{-\frac{1}{2}}+2\sqrt{b}\right).
\end{equation}

By taking care to consider the sign of $\DMLSr{obs}-\DMLSx_n$,
it is straightforward to combine (\ref{EQ13_05_47}), (\ref{EQ13_05_49}),
and (\ref{EQ13_05_52}) to arrive at (\ref{EQ13_05_53}).

\subsection{Proof of Theorem 2}
\label{app_no_flow_degrad}

Similarly to Appendix~\ref{app_no_flow},
we can solve the integration in (\ref{EQ13_05_29}) up to any
$\DMLSt{}$ by solving the following two integrals:
\begin{align}
\label{EQ13_05_35_int}
& \int\limits_0^{\DMLSt{}} \ERF{\frac{a}{{\DMLStau}^\frac{1}{2}}}
d\DMLStau, \\
\label{EQ13_05_37_int}
& \int\limits_0^{\DMLSt{}} {\DMLStau}^\frac{1}{2}
\EXP{-\frac{b}{\DMLStau}}d\DMLStau,
\end{align}
where $a$ could be positive or negative and the latter occurs
only when $\DMLSx_n > \DMLSr{obs}$. To solve (\ref{EQ13_05_35_int}) for
$a > 0$,
we apply the substitution $c = a/{\DMLStau}^\frac{1}{2}$ and
use the indefinite integrals \cite[Eq. 4.1.14]{RefWorks:700}
\begin{equation}
\int \ERF{c}
\frac{dc}{c^3} = \frac{-\ERF{c}}{2c^2}
+ \frac{1}{\sqrt{\pi}}
\int \EXP{-c^2}\frac{dc}{c^2},
\label{EQ13_05_30}
\end{equation}
and \cite[Eq. 2.325.5]{RefWorks:402}
\ifOneCol
\begin{equation}
\int \EXP{-c^n}\frac{dc}{c^m} =
\frac{1}{m-1}\left[\frac{-\EXP{-c^n}}{c^{m-1}}
- n\int \frac{\EXP{-c^n}}{c^{m-n}}dc\right],
\label{EQ13_05_31}
\end{equation}
\else
\begin{multline}
\int \EXP{-c^n}\frac{dc}{c^m} = \\
\frac{1}{m-1}\left[\frac{-\EXP{-c^n}}{c^{m-1}}
- n\int \frac{\EXP{-c^n}}{c^{m-n}}dc\right],
\label{EQ13_05_31}
\end{multline}
\fi
as well as the definition of the error function. It is then
straightforward to show that (\ref{EQ13_05_35_int})
for $a > 0$ is solved as
\begin{equation}
\left(2a^2 + \DMLSt{}\right)\ERF{\frac{a}{\sqrt{\DMLSt{}}}}
+ 2a\sqrt{\frac{\DMLSt{}}{\pi}}\EXP{-\frac{a^2}{\DMLSt{}}}-2a^2.
\label{EQ13_05_33}
\end{equation}

Recalling that $\ERF{\cdot}$ is an odd function,
we solve (\ref{EQ13_05_35_int}) for $a < 0$ as
\begin{equation}
\left(2a^2 + \DMLSt{}\right)\ERF{\frac{a}{\sqrt{\DMLSt{}}}}
+ 2a\sqrt{\frac{\DMLSt{}}{\pi}}\EXP{-\frac{a^2}{\DMLSt{}}}+2a^2.
\label{EQ13_05_34}
\end{equation}

To solve (\ref{EQ13_05_37_int}),
we apply the substitution $c = \sqrt{b\DMLStau}$,
apply (\ref{EQ13_05_31}) twice, and use the definition of the
error function. It is then straightforward to show that
(\ref{EQ13_05_37_int}) is solved as
\begin{equation}
\frac{2}{3}\sqrt{\DMLSt{}}\EXP{-\frac{b}{\DMLSt{}}}
(\DMLSt{} -2b) + \frac{4}{3}b^{\frac{3}{2}}\sqrt{\pi}
\left(1-\ERF{\sqrt{\frac{b}{\DMLSt{}}}}\right).
\label{EQ13_05_37}
\end{equation}

If we take care to consider the sign of $\DMLSr{obs}-\DMLSx_n$, then
we can arrive at (\ref{EQ13_05_38}) by combining (\ref{EQ13_05_33}),
(\ref{EQ13_05_34}), and (\ref{EQ13_05_37}).

\IEEEtriggeratref{31}
\bibliography{../references/nano_ref}

\vspace*{-8cm}

\begin{IEEEbiography}[{\includegraphics[width=1in,height=1.25in,
		clip,keepaspectratio]{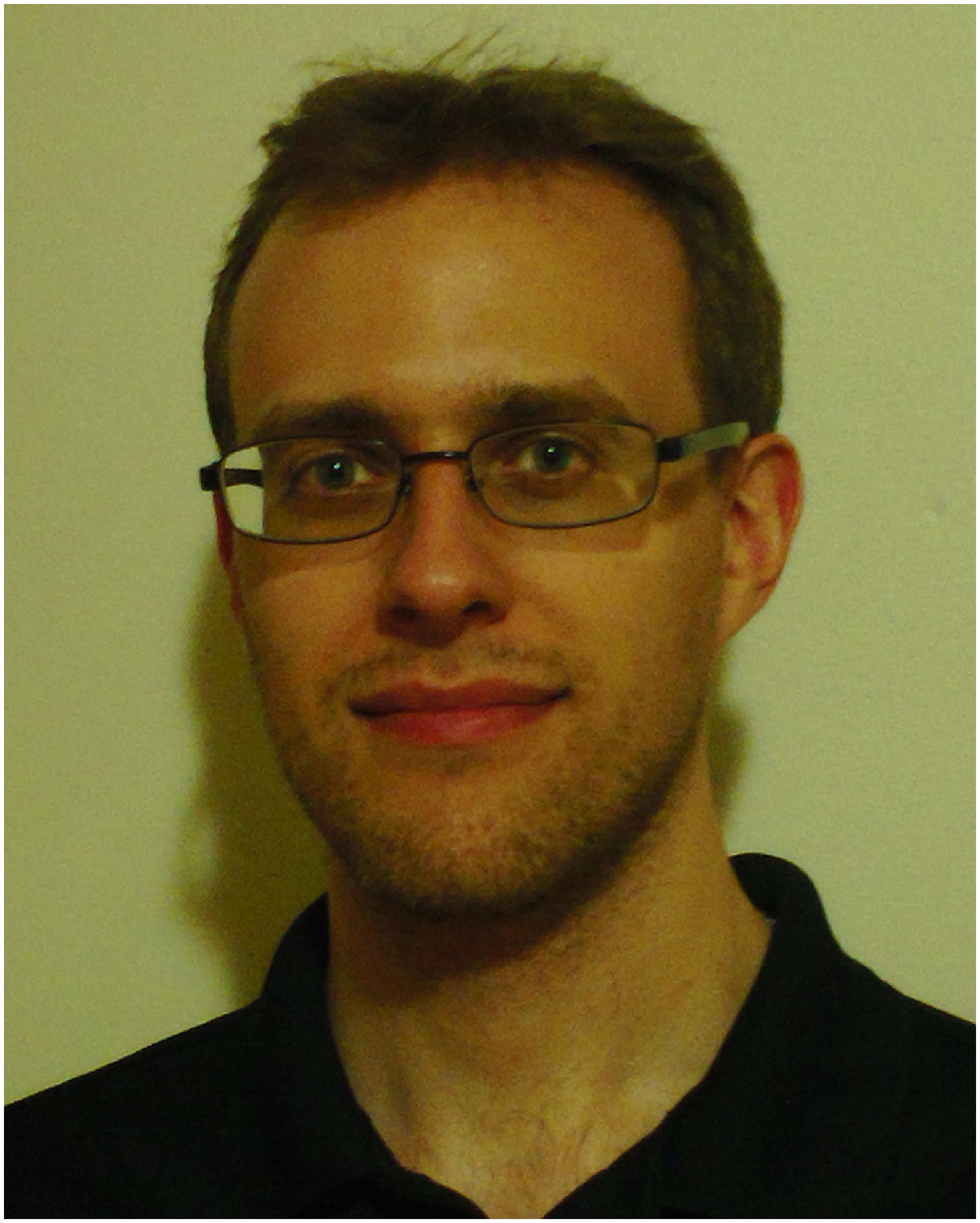}}]{Adam Noel}
	(S'09) received the B.Eng. degree from Memorial University in 2009 and the
	M.A.Sc.
	degree from the University of British Columbia (UBC) in 2011, both in
	electrical engineering.
	He is now a Ph.D. candidate in electrical
	engineering at UBC, and in 2013 was a visiting researcher
	at the Institute for Digital Communications,
	Friedrich-Alexander-Universit\"{a}t Erlangen-N\"{u}rnberg. His research
	interests include wireless communications and how traditional communication
	theory applies to molecular communication.
\end{IEEEbiography}

\vspace*{-8cm}

\begin{IEEEbiography}[{\includegraphics[width=1in,height=1.25in,
		clip,keepaspectratio]{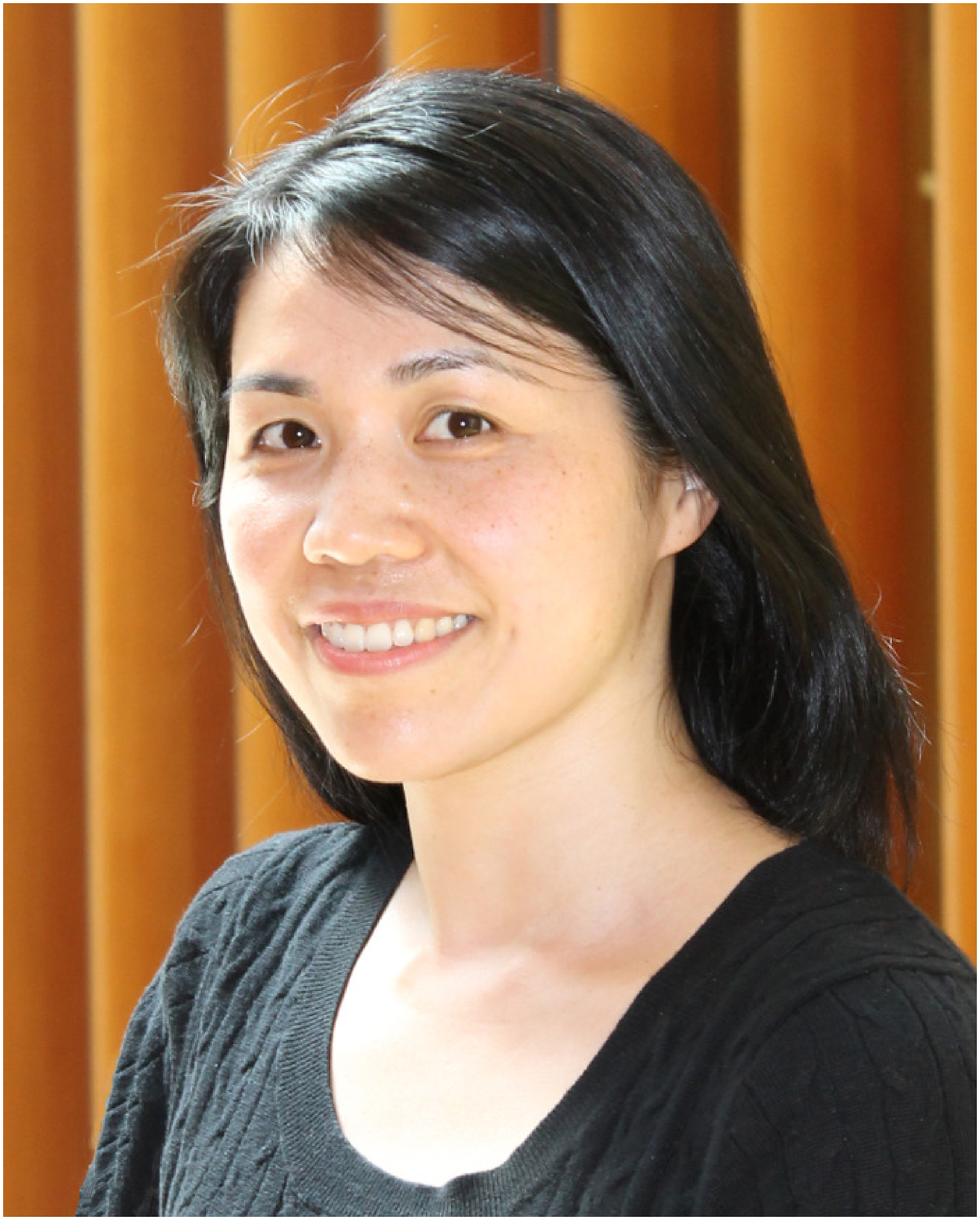}}]{Karen C. Cheung}
	received the B.S. and Ph.D. degrees in bioengineering from the University of
	California, Berkeley, in 1998 and 2002, respectively. From 2002 to 2005, she was
	a postdoctoral researcher at the Ecole Polytechnique Fédérale de Lausanne,
	Lausanne, Switzerland. She is now at the University of British Columbia,
	Vancouver, BC, Canada. Her research interests include lab-on-a-chip systems for
	cell culture and characterization, inkjet printing for tissue engineering, and
	implantable neural interfaces.
\end{IEEEbiography}

\vspace*{-8cm}

\begin{IEEEbiography}[{\includegraphics[width=1in,height=1.25in,
		clip,keepaspectratio]{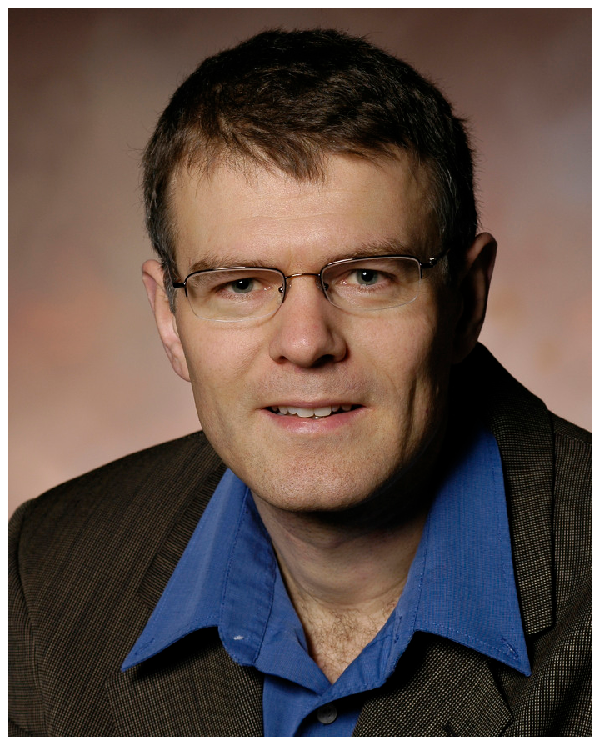}}]{Robert Schober}
	(S'98, M'01, SM'08, F'10)
	received the Diplom (Univ.) and the Ph.D. degrees in electrical engineering from
	the University of Erlangen-Nuremberg in 1997 and 2000, respectively.
	Since May 2002 he has been with the University of British Columbia (UBC),
	Vancouver, Canada, where he is now a Full Professor.
	Since January 2012 he is an Alexander
	von Humboldt Professor and the Chair for Digital Communication at the Friedrich
	Alexander University (FAU), Erlangen, Germany. His research interests fall into
	the broad areas of Communication Theory, Wireless Communications, and
	Statistical Signal Processing.
	He is currently the
	Editor-in-Chief of the IEEE Transactions on Communications.
	
\end{IEEEbiography}

\ifOneCol
	
	\tableOverview{H}{1.8cm}{1.5cm}{1cm}{2.5cm}
	
	\tableNoise{H}
	
	\tableParam{H}
			
	\tableDMLS{H}
	
	\figNoiseNoVK{H}{0.75}
	
	\figNoiseKNoV{!tb}{0.7}
	
	\figNoiseK{!tb}{0.7}
	
	\figNoiseV{!tb}{0.7}
	
	\figNoiseVFar{!tb}{0.7}
	
	\figInterferer{!tb}{0.7}
	
	\figISI{!tb}{0.7}
\fi

\end{document}